\documentclass[aps,pra,twocolumn,showpacs,nofootinbib,superscriptaddress,groupedaddress,a4paper]{revtex4-1}
\usepackage{amsmath}
\usepackage{epsfig}
\usepackage{graphicx}
\usepackage{color}
\usepackage{amssymb}
\usepackage{epstopdf}
\usepackage[utf8]{inputenc}
\usepackage[T1]{fontenc}
\usepackage[colorlinks=True, linkcolor=blue, citecolor=blue]{hyperref}
\usepackage{ulem}
\usepackage{lineno}
\begin{document}
\title{
Faraday waves on a bubble Bose-Einstein condensed binary mixture}
\author{Leonardo Brito$^{1}$}\thanks{brito.phys@gmail.com}
\author{Lauro Tomio$^{2,3}$}\thanks{lauro.tomio@gmail.com}
\author{Arnaldo Gammal$^1$}\thanks{gammal@if.usp.br}
\affiliation{$^{1}$Instituto de F\'{i}sica, Universidade de S\~{a}o Paulo, 05508-090 
S\~{a}o Paulo, Brazil.\\
$^{2}$Instituto de F\'isica Te\'orica, Universidade Estadual Paulista, 01156-970 
S\~ao Paulo, SP, Brazil.\\
$^{3}$Centro Internacional de Física, Instituto de Física, Universidade de Brasília, 70910-900 Brasília, DF, Brazil.}
\date{\today}
\begin{abstract}
By studying the dynamic stability of Bose-Einstein condensed binary mixtures trapped on 
the surface of an ideal two-dimensional spherical bubble, we show how the Rabi coupling
between the species can 
modulate the interactions leading to parametric resonances.  
In this spherical geometry, the discrete unstable angular modes drive both 
phase separations and spatial patterns, with Faraday waves emerging and coexisting 
with an immiscible phase. Noticeable is the fact that,
in the context of discrete kinetic energy spectrum, the only parameters to drive the emergence of
Faraday waves are the $s-wave$ contact interactions and the Rabi coupling. 
Once analytical solutions for population dynamics are obtained, the 
stability of homogeneous miscible species is investigated through Bogoliubov-de Gennes 
and Floquet methods, with predictions being analysed by full numerical solutions applied to
the corresponding time-dependent coupled formalism.
\end{abstract}
\maketitle
\date{\today}
%\pacs{0um3.75.Lm, 67.85.De}
%\keywords{Bose-Einstein condensates, Rabi coupling, bubble trap,Bose-Bose mixtures,
%bubble dynamics}
\section{Introduction}\label{sec1}
Spatial pattern formations can be observed in different branches of physics, whenever 
describing nonlinear wave propagations such as in fluids outside equilibrium and 
nonlinear optics~\cite{1993Cross}. Indeed, surface wave excitations, 
which appear as patterns on liquids inside a vibrating receptacle, were first 
noticed and described by Faraday in 1831~\cite{1831Faraday}, following his famous 
experiments on the formation of patterns in vibrating surfaces. 
The past few-decades progress achieved in reaching near-zero temperatures, allowing
feasibility of Bose-Einstein condensates (BECs) in ultracold
gases~\cite{1995Anderson,1995Davis,1995Bradley}, together with new advanced
techniques to control particle interactions, have opened new ways to explore and 
investigate how some well-known classical phenomena can stand and be realized in 
atomic trapped quantum fluids~\cite{1999Dalfovo,2016Stringari}. 
In atomic gases, spatial patterns can be led by parametric modulations, with 
the emergence of Faraday waves been reported in several experimental and theoretical 
investigations~\cite{
2002Staliunas,2007Engels,2010Nath,2016Sudharsan,2016Abdullaev,2019Abdullaev,2018Smits,2019Nguyen,
2020Zhang,2020Maity,2021Kwon}. 
Particularly concerned to theoretical activities 
on two-dimensional (2D) parametric instabilities in quantum gases, we have the recent work by 
Fujii et al.~\cite{2023Fujii}, in which past and present studies can be traced from the references therein. 
With Faraday patterns in two-component superfluid, quite recently
we had a report in~\cite{2022Cominotti} on the observation of massless and massive collective excitations.
Time-dependent modulations in trap potential or scattering length manipulations 
via Feshbach mechanisms~\cite{1999Feshbach} are able to drive systems to target
excited states \cite{2020Brito}, induce time-crystal formations~\cite{2019Zhu}, 
manipulate population dynamics~\cite{1998Grifoni,2000Williams}, as well as 
explore the Bardeen-Cooper-Schrieffer and BEC crossover, as recently reported
in~\cite{2022He} for bubble-trapped two-component atomic Fermi superfluid. 
On the actual possible technological applications related to manipulations of 
ultracold atoms in matter waves, the so-called atomtronics, we have a recent review 
in~\cite{2022Amico}. In binary dipolar quantum gases, it was also demonstrated 
in~\cite{2020Turmanov} the possibility to create persistent density waves, with 
Faraday instabilities generated by the population imbalance between the two hyperfine
states.
Another interesting way to induce Faraday waves was proposed in \cite{2019Chen}, 
considering interactions effectively modulated by the
Rabi coupling between states. An effective 
interaction actually leads the dynamics\cite{2003Jenkins,2019Shibata}, being able to 
trigger parametric resonances~\cite{1960Landau}.
Also applied by Raman-induced spin-orbit coupling~\cite{2022Zhang}, this approach 
suggests some advantages in dealing with condensate mixtures. 

Currently, condensate mixtures can be performed with the same atomic species initially 
set into different hyperfine states~\cite{1997Myatt}, but also can be handled with 
different atomic species \cite{2002Modugno}. It is possible to study how one species 
is affected by the presence of another one~\cite{2021Alon}. We can observe how the 
elementary excitations can grow and induce phase separation depending on the 
interaction parameters~\cite{2010Brtka, 2021Andriati}. An open question is how Faraday 
waves can be achieved in closed geometries, especially when the excitation is led by 
Rabi coupling. We turn our attention to bubble two-dimensional (2D) bubble shells 
in the spherical closed geometry, where the spectrum of elementary excitation is 
discrete, and therefore spatial patterns need very special conditions to be accomplished.
Following previous theoretical investigations considering 2D shell-like
potentials~\cite{2001Zobay,2004Zobay} confining condensates in spherical 
and/or ellipsoidal surfaces, the recent particular interest is concerned to 
microgravity Bose gas experiments performed aboard the International 
Space Station~\cite{2020Aveline,2022Carollo,2023Lundblad}, 
although bubble BEC has not yet been achieved in space.
Looking for alternative closed 2D geometries, earth experiments have being reported
on a shell bubble BEC produced by exploring the immiscibility of two species~\cite{2022Jia}, 
as well as by controlling a quantum gas confined onto a shell-shaped surface~\cite{Guo2022}.
In this context, several related problems can be traced through recent works
~\cite{2023Tononi,2018Padavic,2018Sun,2019Bereta,2019Tononi,2020Moller}, in
which fundamental properties can be found. More specifically, among others,
we have studies on vortex dynamics and stability~\cite{2020Padavic,2021Andriati,2022Caracanhas,2023Saito},
dipole interactions~\cite{2012Adhikari,2020Diniz}, Berezinskii-Kosterlitz-Thouless transition~\cite{2022Tononi}, 
thermodynamics properties~\cite{2019Pretipino,2021Rhyno}, and bubble mixtures~\cite{2022Sturmer}.

Our focus in the present work is on homogeneous condensates coupled by Rabi
oscillations in a 2D spherical hollow shell.
Inspired by results given in~\cite{2019Chen}, we expect that the stability dynamics 
would be strongly affected by the Rabi coupling, triggering unstable modes, 
which can lead to the rising of new spatial patterns, eventually evolving to 
Faraday waves.
In fact, our findings point out that, the Rabi oscillations are 
able to drive the condensates to states where the Faraday pattern coexists 
with an immiscible phase, where unstable modes are identified by the wave number. 
The same unstable angular modes 
which break a condensate into pieces~\cite{2021Andriati} are able to 
induce Faraday-patterns excitations. 
We first study the stability dynamics by a
comparative analysis of the elementary excitations spectrum obtained via the 
Bogoliubov-de Gennes (BdG)~\cite{2016Stringari} method with the Floquet 
approach \cite{1998Grifoni}. 
Next, by performing the full dynamics with the corresponding 
Gross-Pitaevskii (GP) formalism, we observe that the Floquet approach is more 
suitable to study our coupled system than the BdG scheme, because the Floquet method 
takes into account dynamical effects which cannot be assimilated by the BdG approach.
By tuning the Rabi coupling under given conditions, Faraday waves can emerge,
persisting  even in the immiscible phase. 

The next sections are organized as follows.
In Sect.~\ref{sec2}, we present the theoretical model for a Rabi-coupled two 
BEC mixture confined in the surface of a rigid spherical shell.
The stability analyses are provided in Sects.~\ref{sec3} and ~\ref{sec4};
by applying, respectively, the BdG approach to stationary solutions, and 
the Floquet method to homogeneous oscillating solutions. 
In Sect.~\ref{sec5}, the atom-population dynamics is performed by solving
the full GP formalism, in which stability predictions are checked and we can
observe the emergence of Faraday waves. Finally, in Sect.~\ref{sec6}, 
we have our main conclusions with some perspectives. Among the four 
appendices with complementary material, particular attention should be 
given to Appendix~\ref{app_density}, which provides exact analytical
solutions for binary density oscillations in a spherical bubble.

\section{Rabi-coupled BEC mixture on a rigid spherical shell}
\label{sec2}
We consider a binary BEC mixture with two atomic species sharing the same mass $M$, which
can be in two different hyperfine states. Our study is performed by assuming that both 
condensates are trapped on the surface of a rigid spherical shell, aiming to mimic the 
cold-atom bubble experiments currently performed in microgravity environments. 
For that, a reduction of the original three-dimensional (3D) GP coupled equation is 
performed to a corresponding two-dimensional (2D) formalism.
The 2D approximation is reasonable as long as the radial excitations are 
inaccessible regarding the large amount of energy needed for it. This is true 
when the thickness $\delta R$ of the 3D spherical shell of radius $R$ is comparatively
very small $(\delta R \ll R$), as extensively discussed in~\cite{2021Andriati}. We also
stress that our main concern is on the {\it dynamic stability} of the system, 
{\it i.e.}, a context where we are not taking into account {\it energetic instabilities},
which could be triggered by a thermal cloud that is neglected here. 

The condensates can be described in the mean-field approach as a system 
of two coupled GP equations \cite{1999Dalfovo,2016Stringari}, with atoms transferred 
from one state to the other by Rabi oscillations~\cite{2019Chen}, by taking into 
account real Rabi coupling $\Omega$. With the total number of atoms $N$, the coupled
condensates  with respective populations are given by $N_1(t)$ and $N_2(t)$. They 
interact each other through their nonlinear two-body parameters $\overline{g}_{ij}\equiv4\pi\hbar^2{a}_{ij}N/M$ 
($i,j=1,2$), where $a_{jj}$ and $a_{12}=a_{21}$ are, respectively, the intra- and 
inter-species $s$-wave atom-atom scattering lengths.  With this definition, we are assuming the total wave function normalized 
to 1, with each component $j = 1, 2$ normalized to $N_j(t)/N$.
Along this paper, with exception of Appendix~\ref{2D_red+adim}, 
we use dimensionless variables and quantities, by taking the bubble radius $R$ 
as the length unit, with $\hbar^2/(MR^2)$ and $MR^2/\hbar$ being the energy and time 
units, respectively. The dimensional reduction, from 3D to 2D, is detailed in 
the Appendix~\ref{2D_red+adim}, with the adimensionalization being explained at
the end by factoring the energy unit. In this way, we end up with the following 
2D coupled GP formalism describing each wave function 
$\Psi_j\equiv\Psi_{j}(\theta,\phi;t)$ normalized to $N_j$:
\begin{eqnarray}
{\rm i}\partial_t \Psi_{j=1,2}&\!=\!&-\frac{1}{2 \sin\theta}\left[
    {\partial_\theta}\left(\sin\theta\,
    {\partial_\theta}\right)+\frac{1}{\sin\theta}{\partial_\phi^2} 
\right]\Psi_{j}\nonumber\\
&+&\sum_{i=1,2}g_{ji}{\left| \Psi_{i}\right|}^{2}
\Psi_{j}+(-1)^{j}{\mathrm i}\Omega\Psi_{3-j}, \label{eq01}
\end{eqnarray}
where $\theta\in[0,\pi]$ and $\phi\in[0,2\pi]$ are, respectively, the usual 
polar and azimuthal angular positions in the sphere; and, the notation
$\partial_\chi$ is being used for partial derivative of $\chi$. Also, 
represented in the 2D spherical surface, 
$g_{ij}=\frac{\sqrt{8\pi}a_{ij}N}{\delta R}$ (with $\delta R$ being the 
thickness of the bubble shell)
are the nonlinear parameters for the inter- and intra-species interactions, 
derived in Appendix~\ref{2D_red+adim} from the original 3D formalism
after factoring the energy unit~\footnote{The constant factor in the definition
of $g_{ij}$ is model dependent,
with the given factor obtained by assuming a Dirac-delta-like radial Gaussian 
function with center in $R$ and width $\delta R\ll R$.}.
Our purpose is to study the stability of a miscible homogeneous system under
different conditions, considering stationary as well as time-oscillating
solutions. 

The spatial part of \eqref{eq01} is directly proportional do the square of 
the angular momentum $\mathbf{L}$, given by ${\bf L}^{2}/(2\hbar^2)$, 
with exact discrete $\ell-$state eigenvalues $\epsilon_\ell=\ell(\ell+1)/2$, 
corresponding to spherical harmonics eigenfunctions $Y_{\ell,m}(\theta,\phi)$. 
Therefore, it is appropriate to redefine the component wave functions for specific $\ell$ states as $\Psi_j(\theta,\phi;t)\equiv\psi_{j,\ell}(t)Y_{\ell,m}(\theta,\phi)$, such that, as $-\ell\le m\le \ell$, we can have $2\ell+1$ states with the
same eigenvalue $\epsilon_\ell$.
For convenience, the explicit time and $\ell$ labels will be removed within the redefinition $\psi_{j}\equiv \psi_{j,\ell}(t)$.
In order to verify the time-dependent oscillatory behavior of $\psi_j$, 
let us assume the simplest case with $\Omega=0$, $N_j=N/2$, $g_{jj}=g$ 
and $g_{21}=g_{12}$, with the stationary-state wave functions identical for 
both components,  
\begin{equation}
    \psi^{(s)}_{j}=\sqrt{\frac{1}{8\pi}}
    \;\exp\left[{-{\mathrm i}\displaystyle\left(
    \frac{g+g_{12}}{8\pi}
    \right)}t\right],   
    \label{eq02}
\end{equation}
where $(g+g_{12})/{(8\pi)}$ is the chemical potential. 
This stationary case can be easily extended to homogeneous periodic solutions, 
for $g_{12}=g$ and $\Omega\ne 0$.
By including a time-dependent oscillating factor in the normalization, 
implying in periodic exchange of atoms between the species, 
the coupled wave functions are expressed by
\begin{eqnarray}
\left(\begin{array}{l}
\psi_1^{(0)}\\
\psi_2^{(0)}
\end{array}\right)
=
\sqrt{\frac{1}{4\pi}}
\left(\begin{array}{l}
\cos\left(\Omega t+\frac{\pi}4\right)\\
\sin\left(\Omega t+\frac{\pi}4\right)
\end{array}\right)\exp\left[{{-{\rm i}\left(\displaystyle\frac{g }{4\pi}\right)t}}\right]
\label{eq03}.
\end{eqnarray}
The period of oscillations is given by
$2\pi/\Omega$, which does not depend on the interactions, being twice the
period for the densities $T_0\equiv\pi/\Omega$.
For $\Omega\ne 0$ and $g_{12} \ne g$, 
a more general periodic solution of \eqref{eq01} can be derived, with
\begin{equation}
    \psi_{j}(t)=f_{j}(t)\exp{(-{\mathrm i}\gamma_j t)},
    \label{eq04}
\end{equation}
where $f_{j}(t)$ are periodic complex functions (satisfying $|f_{j}(t+T)|^2=|f_{j}(t)|^2$ 
for a period $T$), with $\gamma_j$ being real and time-independent phases.
An equivalent approach is to write $f_j(t)$ as real functions with time-dependent phases 
to be determined. By assuming both states equally populated at $t=0$, 
we have $|\psi_{j}(0)|^2={1/(8\pi)}$.
The atom-number ratios of the two-atomic species, given by $N_j(t)/N=4\pi|\psi_j(t)|^2$, 
oscillate periodically within a cycle given by the period $T$ and amplitude
${\cal A}$ (maximum exchange number-ratio of particles), which in 
general are functions of the Rabi parameter $\Omega$ and nonlinear
interactions $g_{ij}$. As shown for $g_{12}=g$, $T=T_0=\pi/\Omega$ does not 
depend on the interactions, with each density $|\psi^{(0)}_j(t)|^2$ oscillating 
between zero and $1/4\pi$. 
However, it can be shown that the period decreases as $|g-g_{12}|$ increases, 
with the maximum occurring when $g_{12}=g$~\cite{2019Chen}.

By considering Eq.~\eqref{eq04} with $\Delta g \equiv (g_{12}-g)\ne 0$ 
and $\Omega\ne 0$, as shown in Appendix~\ref{app_density}, 
one can obtain the general solution, with time-dependent phases 
$\gamma_j$ not identical for both species
[as verified in \eqref{gamma2}, where $\gamma_j$ is replaced by $\bar{\gamma}_j(t)$,
the phases will depend on the respective densities $|\psi_j|^2$], .
However, for the following stability analyses, we can consider they are
carrying only the identical constant part, $\gamma_1=\gamma_2\equiv\gamma_0=(g+g_{12})/(8\pi)$, such that the 
time-dependent parcel of the phases are retained by the complex functions $f_j$, 
which in this case are satisfying 
{\small\begin{eqnarray}
&&\partial_{tt}f_{j}+\left[\Omega^2+(\Delta g)^2
\left(\frac{|f_2|^2-|f_1|^2}{2}
\right)^2\right] f_j=0. \label{eq05}
\end{eqnarray}
}The derivation of the above, obtained from \eqref{eq01} with \eqref{eq04}, 
follows analogously as shown in Appendix~\ref{app_density}, from 
\eqref{gp2} to \eqref{w2t}. For $g_{12}=g$, this equation is identified with the harmonic oscillator equation 
having the frequency given by the Rabi parameter. The atom-ratio difference, 
which is defined by $\nu(t)\equiv \frac{N_2-N_1}{N}$  $=4\pi(|f_2|^2-|f_1|^2)$ $=4\pi
\left(|\psi_2|^2-|\psi_1|^2\right)$,
satisfies the undamped unforced Duffing oscillator motion described by~\cite{2014Salas,2016Belendez},
\begin{eqnarray}
&&\partial_{tt}\nu+4\left[\Omega^2+\frac{1}{2}
\left(\frac{\Delta g}{8\pi}\right)^2\nu^2\right]\nu=0, \label{duffing}
\end{eqnarray}
which has exact solutions given by Jacobi elliptic periodic functions, for which 
the period is expressed by
{\small\begin{eqnarray}
T_K(\Omega,\alpha)&=&
\frac{2}{\sqrt{{\Omega^2}+\alpha^2}}K\left(\frac{-\alpha^2}{\Omega^2+\alpha^2}\right), \\
\;{\rm with}\;
\alpha &\equiv& {\left(\frac{{\cal A}\,\Delta g}{8\pi}\right)}, \label{period-ell}
\end{eqnarray}
}where ${\cal A}$ is the amplitude of the density difference $\nu$ oscillations, 
with $K(x)$ being the first-kind Jacobi elliptic function~\cite{2021Whittaker}. 
For details, see Appendix~\ref{app_density}), where it was shown that this oscillating 
period is exact and can be obtained even before the explicit form of $\nu(t)$ is
obtained. Here, for a convenient resemblance with the harmonic oscillator
sinusoidal form, we assume $\nu(t)$ identified with
{\small\begin{eqnarray} 
\nu(t)&=&{\cal A}
\sin\left[2t
\sqrt{\Omega^2+\frac{\alpha^2}{2}\left(\frac{\nu_A}{\cal A}\right)^2}\right],
\label{atom-diff}\end{eqnarray}
}where, for the moment, $\cal A$ and $\nu_A$ are parameters, which can be 
obtained from the exact solution of the Duffing equation.
Within the present normalization of the coupled equation, 
${\cal A}\le 1$, being 1 for $(\Delta g)=0$. As derived in 
the Appendix~\ref{app_density}, constrained by the periodic conditions, 
$\cal A$ is a function of the ratio $(\Delta g)/\Omega$ 
given by
\begin{eqnarray}
\frac{\cal A}{8\pi}={\sqrt{2}\frac{\Omega}{\Delta g}}
\left[\sqrt{1+\left(\frac{\Delta g}{8\pi\Omega}\right)^2}-1\right]^{1/2}.
\label{amplitude}\end{eqnarray}
With the atom-ratio difference expressed by \eqref{atom-diff}, 
the exchange oscillating time interval (from ${\cal A}$ to
$-{\cal A}$), is one-half of the density period, given by
\begin{equation}
T=\frac{\pi}{\sqrt{\Omega^2+\frac{\alpha^2}{2}\left(\frac{\nu_A}{\cal A}\right)^2}}
\label{period},
\end{equation}
which has an exact agreement with \eqref{period-ell}, for 
$\alpha=0$ ($\Delta g=0$), $K(0)={\pi}/2$. 
By matching \eqref{period} with \eqref{period-ell} in the other extreme, 
$\Omega=0$ (the stationary limit), we obtain  
\begin{eqnarray}
    \left(\frac{\nu_A}{\cal A}\right)=\left(\frac{\pi}{ \sqrt{2} K(-1) }
    \right)=1.6945.
\end{eqnarray}
\begin{figure}[h]
    \centering
    \includegraphics[scale=0.22]{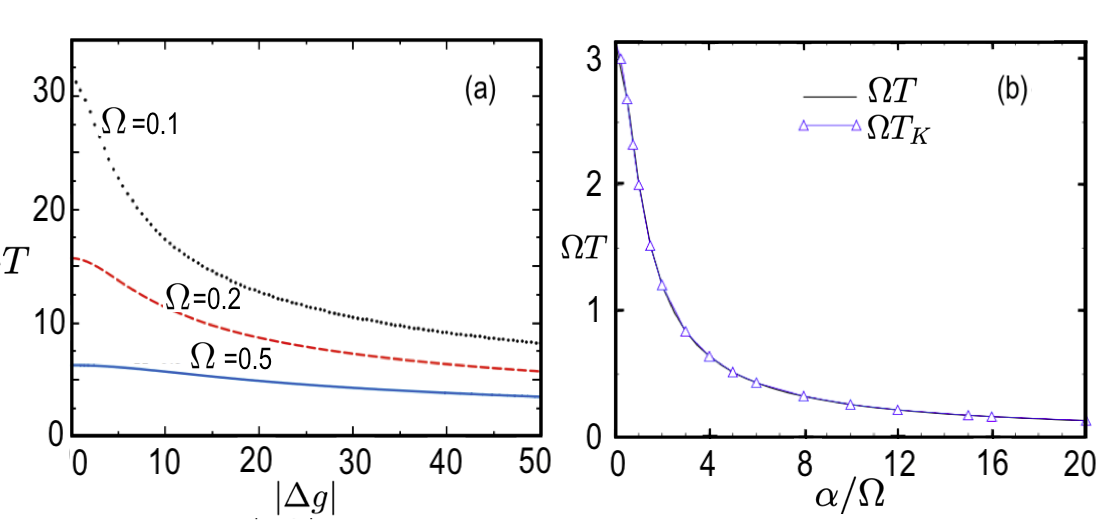}
\vspace{-0.2cm}
\caption{(Color online) In panel (a), the density oscillating period $T$ is given 
as a function of the absolute difference of the interaction parameters $|\Delta g|
$, for three different Rabi couplings $\Omega$, as indicated. In panel (b), it is 
shown the perfect agreement between analytical expressions for the {\it Duffing}
period $T_K$ (empty-triangles) and $T$ (solid-line), respectively 
multiplied by $\Omega$, given by \eqref{period-ell} and \eqref{period},
with $\alpha=\frac{{\cal A}\Delta g}{8\pi}$.
Within defined units, all quantities are presented as dimensionless.
}
\label{fig01}
\end{figure}
\begin{figure}[h]
\includegraphics[scale=0.25]{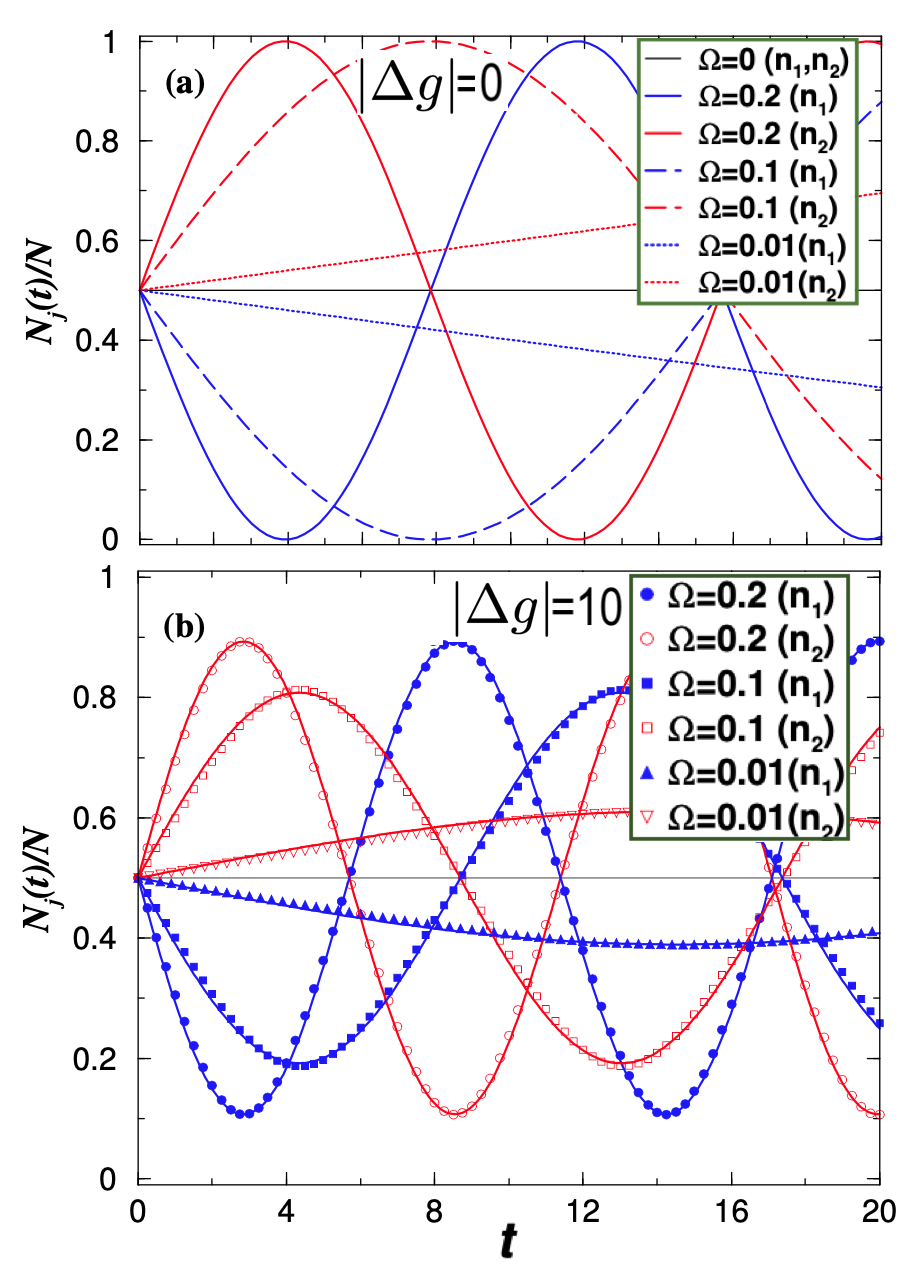}
\vspace{-0.2cm}
\caption{(Color online) Time-evolution of the atom-number ratio, 
$N_j(t)/N\equiv 
4\pi|\psi_{j}(t)|^{2}\equiv 4\pi n_j$, with initial 
condition $N_j(0)=N/2$, for given Rabi couplings $\Omega$. 
In both panels, (a) for $|\Delta g|=0$, 
and (b) for  $|\Delta g|=10$, 
the initially decreasing (increasing) lines refer to 
species 1 (species 2) [horizontal line for $\Omega=0$]. 
In (b), the full-numerical solutions (legend box),
are matching with corresponding solid lines, given by 
\eqref{atom-diff}, for $({\cal A},\Omega) =(0.7873,\;0.2)$, 
$(0.6261,\;0.1)$, and 
$(0.2214,\;0.01)$, 
with $(\nu_A/{\cal A})=\pi/\left(\sqrt{2} K(-1)\right)=1.6945$.
Within defined units, all quantities are presented as dimensionless.
}
\label{fig02}
\end{figure}
In Fig.~\ref{fig01}, where we are displaying exact numerical results for the
dependence of the oscillating period $T$ on the interaction difference
$\Delta g$ for a few values of the Rabi frequency $\Omega$ [panel (a)], 
we also show the perfect agreement between the expressions \eqref{period} and 
\eqref{period-ell} in panel (b).
In panel (a), one can also notice explicitly how the oscillation period 
diminishes as the Rabi coupling increases. 

To illustrate the density behavior, when considering different 
Rabi couplings 
and interactions, we also present two panels in Fig.~\ref{fig02}, with 
few samples of full-numerical solutions for the corresponding
density oscillations, together with close-approximated analytical 
solutions obtained by considering \eqref{eq05}, as detailed in 
Appendix~\ref{app_density}.
Panel (a) shows the behavior of the densities for the particular 
cases with $g_{12}=g$ when the two components follow the 
simple analytical expressions 
\eqref{eq03}. In contrast, panel (b) illustrates the behavior of a
more general case with $g_{12}\neq g$, according to Eq.~\eqref{eq04},
when the solutions are deviating from the simple sinusoidal form. 
For a given Rabi parameter $\Omega$, as the differences
between the interactions ($|g-g_{12}|$) increase, the number of particles
being exchanged (represented by the corresponding amplitudes) decreases, 
oscillating within a smaller interval. Correspondingly,
also noticeable in Figs.~\ref{fig01}-\ref{fig02}, 
is the effect of symmetry breaking; when we broke the perfect balance 
between intra- and inter-species interactions the 
inter-condensate atom exchange frequency increases.

The perfect agreement between the expressions \eqref{period} and 
\eqref{period-ell} is shown in Fig.~\ref{fig02}{(b)}, which implies that
\eqref{atom-diff} is a close approximation to the exact solution
of the Duffing equation. This fact is confirmed by some sample results
shown in Fig.~\ref{fig02}{(b)}, where analytical results are 
compared with full numerical ones. 
From \eqref{atom-diff} both densities can be written as 
{\small\begin{eqnarray} 
|f_j(t)|^2
&=&\frac{1}{8\pi }+(-)^j
\frac{\cal A}{8\pi }
\sin\left[2t
\sqrt{\Omega^2+\frac{\alpha^2}{2}\left(\frac{\nu_A}{\cal A}\right)^2}
\right], \nonumber \\
&{\longrightarrow}&\frac{1}{4\pi}\left\{ \begin{array}{c}
\cos^2(\Omega t)\\
\sin^2(\Omega t)\end{array}\right.\;{\rm for}
\;\Delta g\to 0\; ({\cal A}\to 1).
\label{eq06}
\end{eqnarray}
}As the coupled system is normalized to one, with one density 
orthogonal to the other,  
the extremes for the difference are $\pm{\cal A}$, 
which happens when one of the species is at the maximum with the 
other at the minimum. 
When $\Omega\to 0$ (or $\Omega\ll |\Delta g|/(8\pi)$), 
we have the other extreme, with~\eqref{eq06} satisfying the stationary
case \eqref{eq02}, 
with both densities being identical, $|f_j|^2=1/(8\pi)$.
As $|\Delta g|$ increases, the periodic atom exchange between the 
coupled condensates decreases till 
reaching the stationary limit.

With respect to the Rabi frequency $\Omega$, 
more time is needed for an oscillating solution to complete each cycle with 
lower values of $\Omega$ than for higher ones.
As verified in Fig.~\ref{fig02}, for the initial time interval, lower
frequencies provide almost linear behaviors (increasing or decreasing) with time 
when compared with the corresponding behavior obtained with higher frequencies. 
So, at short times, when the Rabi coupling is weak ($\Omega\rightarrow 0$), 
stationary solutions and oscillating ones are likely to be the same. This is no 
longer true for strong coupling. 

\section{Bogoliubov-de Gennes stability analysis}\label{sec3}
The role of the Rabi coupling $\Omega$ on stationary solutions 
(\ref{eq02}) is studied in this section by performing a dynamic 
stability analysis, using the BdG method~\cite{2010Brtka,2021Andriati}. 
Within this approach, small amplitude oscillations are considered around the 
uniform stationary solution \eqref{eq03}. With the perturbations 
being eigenfunctions of the kinetic energy operator, we can express the 
perturbed wave functions by $\ell-$angular-mode oscillations, in terms of 
the spherical harmonics $Y_{\ell, m}(\theta,\phi)$:
{\small
\begin{eqnarray}
\psi_{j,\ell}^{(s)}(\theta,\phi;t)=
\Big\{ \sqrt{\frac{1}{8\pi}}
&+& u_{j,\ell}^{(s)}Y_{\ell,m}(\theta,\phi){\mathrm e}^{-{\mathrm i}
    \omega_{\ell}t} 
\nonumber\\&+&
    {v}_{j,\ell}^{(s)*}{Y}_{\ell,m}^{*}(\theta,\phi)
    {\mathrm e}^{{\mathrm i}{\omega}_{\ell}^{*}t}
\Big\}
    \mathrm{e}^{-{\mathrm i}\mu t}, \label{pert}
\end{eqnarray}
}where $u_{j,\ell}^{(s)}$ and $v_{j,\ell}^{(s)}$ are complex parameters
to be determined. The spectral solutions are given by $\omega_\ell$, with $\ell$ being specific angular 
mode oscillations. Therefore, all the perturbation terms of \eqref{pert} are exact solutions of the 
linear part of \eqref{eq01}~\cite{2021Andriati},
with eigenvalues $\epsilon_\ell\equiv \ell(\ell+1)/2$. The particular simplified symmetric form of 
Eq.~\eqref{eq01} allows us to assume perturbations with no dependence on the azimuthal mode excitation, 
given by $m$ (an integer running from $-\ell$ to $+\ell$), which can be 
arbitrarily chosen.
Therefore, in the exponential factors of Eq.~\eqref{pert}, the frequency parameters 
$\omega_\ell$  are excitation modes that carry only the angular momentum 
index $\ell$. They are in general complex numbers, with non-zero imaginary parts 
when the system becomes dynamically unstable. By initially assuming they are 
real numbers, we are considering parameters such that the system is in a 
stable configuration. As we vary these parameters,
for some specific modes of oscillation the system becomes unstable, 
acquiring non-zero imaginary parts.

By inserting the perturbation \eqref{pert} into the Eq.~(\ref{eq01}), neglecting the second 
and higher-order amplitude terms, we obtain the corresponding BdG matrix equation,
\begin{eqnarray}
\left[{\mathbf M^{(s)}}-\omega_{\ell}\right]{\mathbf u}_{\ell}^{(s)}=0,\label{BdGMatrix}
\end{eqnarray}
where $\bf M^{(s)}$ contains the model parameters
$g_{ij}$ and $\Omega$, 
{\small\begin{eqnarray}{\mathbf M^{(s)}}=
\begin{bmatrix}
\epsilon_{\ell}+\frac{g}{8\pi}&\frac{g}{8\pi}&\frac{g_{12}}{8\pi}-{\mathrm i}\Omega&\frac{g_{12}}{8\pi}\\
-\frac{g}{8\pi}&-\epsilon_{\ell}-\frac{g}{8\pi}&-\frac{g_{12}}{8\pi}&-\frac{g_{12}}{8\pi}-{\mathrm i}\Omega\\
\frac{g_{12}}{8\pi}+{\mathrm i}\Omega&\frac{g_{12}}{8\pi}&\epsilon_{\ell}+\frac{g}{8\pi}&\frac{g}{8\pi}\\
-\frac{g_{12}}{8\pi}&-\frac{g_{12}}{8\pi}+{\mathrm i}\Omega&-\frac{g}{8\pi}&-\epsilon_{\ell}-\frac{g}{8\pi}
\end{bmatrix},
\end{eqnarray}
}and ${\mathbf u}_{\ell}^{(s)}$ is the 
column vector defined by the perturbed amplitudes in \eqref{pert}, with transpose 
$\Big[u_{1,\ell}^{(s)} \ v_{1,\ell}^{(s)} \ u_{2,\ell}^{(s)} \
v_{2,\ell}^{(s)}\Big]^{\rm T}$.
By solving the corresponding determinant, 
four possible solutions for each $\ell$ mode
are obtained, given by
{\small
\begin{eqnarray}\hspace{-0.5cm}
    \omega_{\ell,\pm}^2\!\!&=&\!\!
    \left(\epsilon_{\ell}^2+\frac{\epsilon_{\ell}\;g}{4\pi}\right)+\Omega^2 
    \pm2\sqrt{\left(\epsilon_{\ell}^2+\frac{\epsilon_{\ell}\;g}{4\pi}\right)\Omega^2+ \frac{\epsilon_{\ell}^2\;g_{12}^2}{(8\pi)^2}}. \label{bdg_spc}
\end{eqnarray}}
However, two of them with opposite overall signs are redundant as they correspond to  
exchanging signals in the original definitions, such that only the positive ones will
be considered.
The system is said to be {\it dynamically stable} if these frequencies are real: 
$\mathrm{Im}(\omega_{\ell,\pm})= 0$;
becoming {\it dynamically unstable} when some of the solutions became complex:
$\mathrm{Im}(\omega_{\ell,\pm})\neq 0$.
\begin{figure}[h]
    \centering
    \includegraphics[scale=0.38]{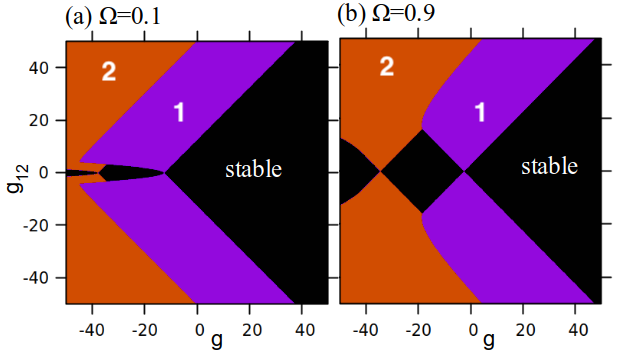}
\vspace{-0.2cm}
\caption{(Color online) BdG stability diagrams for the interaction parameters 
$g_{12}$ vs $g$, as given by~\eqref{bdg_spc}. Stable regions
[Im($\omega_{\ell,\pm}$) = 0] are represented in black, with unstable
$\ell-$modes [Im$(\omega_{\ell,\pm})$ $\neq0$] (with $\ell$ values indicated inside
the regions) are in colors [violet for $\ell=1$ and orange for $\ell=2$]. 
The Rabi coupling $\Omega=0.1, 0.9$ are indicated at the top of the respective panels. Within defined units, all quantities are presented as dimensionless.}
\label{fig03}
\end{figure}
In Fig.~\ref{fig03}, we present a 
BdG stability diagram, which depicts how large is the stable region as one varies  
the interactions. By comparing the panels \ref{fig03}{(a)} 
and \ref{fig03}{(b)}, we verify
that the Rabi strength has an important role in rising new stability regions. 

\section{Floquet stability analysis}\label{sec4}
Let us now consider the evolution of the system given by the homogeneous oscillating 
solutions \eqref{eq04}, under small amplitude oscillations,
in order to better take into account the role of Rabi coupling in the dynamics. 
For that, we can study the solutions dynamically by the time-dependent 
Floquet method~\cite{2019Chen,2022Zhang}, using small amplitude 
oscillations.
For the present Floquet stability analysis, we are considering that
the phases of the wave-function component $\psi_j(\theta,\phi;t)$ 
have a common time-independent part given by  
$\gamma_j=\gamma=(g+g_{12})/(8\pi)$, with $f_{j}(t)$ being complex,
carrying any other relevant part of the phases. 
Another equivalent approach, with time-dependent phases $\gamma_j$ 
and real $f_j(t)$, is detailed in the Appendix~\ref{app_density}. 
In this case, by applying small amplitude oscillations in~\eqref{eq04}, we have  
{\small\begin{eqnarray}
    \psi_{j}(\theta,\phi;t)&=&f_{j}(t)\exp\left({-{\mathrm i}
    \gamma t}\right)\label{period_sol_pert}\\
    &+&\left[{u}_{j,\ell}Y_{\ell, m}(\theta,\phi)+
    {v}_{j,\ell}^{*}Y^*_{\ell,m}(\theta,\phi)\right]{\rm e}^{-{\rm i}\gamma t}, 
    \nonumber 
\end{eqnarray}
}where the amplitudes ${u}_{j,\ell}\equiv{u}_{j,\ell}(t)$ and
${v}_{j,\ell}\equiv{v}_{j,\ell}(t)$ are periodic time-dependent functions, 
with the same period $T$ as $f_j$, such that $|\psi_j (t + T )|^2 = |\psi_j (t)|^2$. 
By inserting (\ref{period_sol_pert}) into (\ref{eq01}), 
neglecting the second and higher-order terms, we get the following 
matrix equation \cite{2019Chen}:
{\small
\begin{eqnarray}
{\mathrm i}\frac{d}{ dt}\!\!
\left[\begin{array}{c}
{u}_{1,\ell}\\
{v}_{1,\ell}\\ 
{u}_{2,\ell}\\ 
{v}_{2,\ell}
\end{array}\right]\!\!=\!\!
\begin{bmatrix}
{D}_{1}                 && G_{11} && D_{12}    &&G_{12} \\
-G_{11}^{*}              &&-{D}_{1}&&-G_{12}^{*}&&-D_{12}^* \\
D_{12}^*                &&G_{12}  &&{D}_{2}    &&G_{22}\\
-G_{12}^{*}             &&-D_{12} &&-G_{22}^{*}&&-{D}_{2}
\end{bmatrix}\!\!
\left[\begin{array}{c}
{u}_{1,\ell}\\
{v}_{1,\ell}\\ 
{u}_{2,\ell}\\ 
{v}_{2,\ell}
\end{array}\right]\label{floquet_eq}\!\!,
\end{eqnarray}
}where the elements are
${D}_{j}\equiv(\epsilon_{\ell}-\gamma)+2g{|f_{j}|}^{2}+g_{12}{|f_{3-j}|}^{2}$, 
$D_{12}\equiv g_{12}f_{1}f_{2}^{*}-{\rm i}\Omega $, and $G_{ij}\equiv g_{ij}f_{i}f_{j}$
($g_{jj}=g$).
When the system is driven by a periodic time-dependent Hamiltonian, the Floquet theorem 
\cite{1998Grifoni} predicts that the
solutions ${{\mathbf u}}_{\ell}(t)$ can be written as
\begin{equation}
    {\mathbf u}_{\ell}(t)=\exp{({\lambda}_{\ell} t)}{\mathbf p}_{\ell}(t),
\end{equation}
where ${\mathbf p}_{\ell}$ are periodic functions, which in our case
satisfy the same periodicity of the densities. The factor 
$\lambda_{\ell}$ stands for the Floquet exponent.  
From its periodic property at the time $t=T$, 
${\mathbf p}_{\ell}(T)={\mathbf p}_{\ell}(0)$, we obtain
\begin{equation}
    {\mathbf u}_{\ell}(T)=\exp{(\lambda_{\ell} T)}
    {\mathbf p}_{\ell}(0).
    \label{floquet_exp}
\end{equation}
\noindent{\bf Numerical approach:}
 Our approach to performing the Floquet stability analysis relies on an exact 
 numerical calculation of the relevant observables. An analytical approach  to 
 obtain the associated wave functions with their small oscillating amplitudes 
 can only be done at some approximate level (as discussed in Sec.~\ref{sec2} and 
 Appendix~\ref{app_density}). Therefore, for practical purposes, we follow a 
 method similar to~\cite{2019Chen}, by integrating the Eq.~\eqref{floquet_eq} using 
 a fourth-order Runge-Kutta method (RK4) from $t=0$  to $t=T$ (a complete period),
 assuming four different initial conditions for the amplitudes, which are 
 ${\mathbf u}_{\ell}(0)=$ $[1 \ 0 \ 0 \ 0]^{\rm T}$, 
 $[0 \ 1 \ 0 \ 0]^{\rm T}$, $[0 \ 0 \ 1 \ 0]^{\rm T}$, 
 and $[0 \ 0 \ 0 \ 1]^{\rm T}$, being the four column 
 vectors of a matrix ${\mathbf F}$, 
that at $t=0$ is the identity matrix ${\mathbf F} (t=0)= 
 [{\mathbf u}_{\ell}^{(1)}(0) \ {\mathbf u}_{\ell}^{(2)}(0) \ 
 {\mathbf u}_{\ell}^{(3)}(0) \ {\mathbf u}_{\ell}^{(4)}(0)]={\mathbf I}$.
  Separately, each initial condition vector will correspond to a different vector 
  at $t=T$, that will define the evolved matrix as ${\mathbf F}(t=T)=
 [{\mathbf u}_{\ell}^{(1)}(T) \ {\mathbf u}_{\ell}^{(2)}(T) \ 
 {\mathbf u}_{\ell}^{(3)}(T) \ {\mathbf u}_{\ell}^{(4)}(T)]$.
 By considering the eigenvalues of ${\mathbf F}$ given by $F_{\lambda_\ell}$, 
 and once identified the matrix with \eqref{floquet_exp},
 we are able to obtain the Floquet exponent as 
 $\lambda_{\ell}=\ln(F_{\lambda_\ell})/T$. 
 If the system evolves to the time $t=T$ with nonzero real part in the full spectrum
 having $(\lambda_{\ell}^{R})>0$, it implies that solution ${\mathbf u}_{\ell}$ is 
 growing exponentially with $t$ for that specific mode $\ell$, being no longer 
 stable. In other words, the uniform oscillating system is dynamically unstable 
 under that $\ell$ orbital excitation. 
 By using this approach, we are also able to study how far the BdG approach 
 returns consistent results. As known, the BdG stability method is suitable for
 stationary solutions. However, when the system is under fast Rabi oscillations, 
 the Floquet approach can give us more reliable results, which can indicate the 
 level of disagreement with results obtained by the BdG method.  
 In the Appendix \ref{app_comparison}, we provide a more detailed comparison between
 both methods.
 \begin{figure}[h]
    \centering
    \includegraphics[scale=0.35]{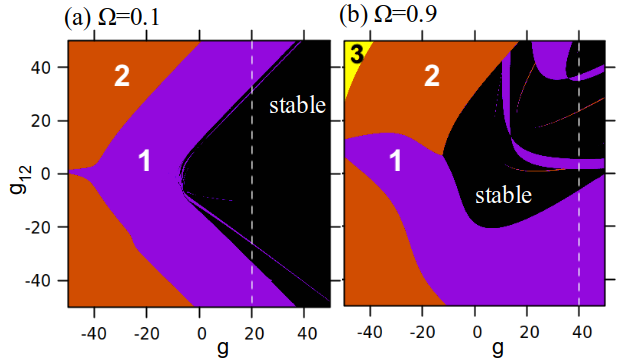}   
\vspace{-0.2cm}
\caption{(Color online) Floquet stability diagrams for constant couplings,
$\Omega=0.1$ (a)  and $\Omega=0.9$ (b), parametrized by 
the interactions $g_{12}$ vs $g$, determined by $(\lambda_\ell^R)_{\rm max}$ 
[See \eqref{floquet_eq}-\eqref{floquet_exp}]. 
The stable regions [$(\lambda_\ell^R)_{\rm max}\le 0$] are in black,
with unstable ones [$(\lambda_\ell^R)_{\rm max}> 0$] having the  
$\ell-$mode given in colors [violet ($\ell=1$), orange ($\ell=2$) and yellow
($\ell=3$)]. The dashed lines, at $g=20$ (a) and $g=40$ (b) refer to results
presented, respectively, in (a) and (b) of Fig.~\ref{fig05}. 
Within defined units, all quantities are presented as dimensionless.
}
\label{fig04}
\end{figure}
\begin{figure}[h]
    \centering
       \includegraphics[scale=0.48]{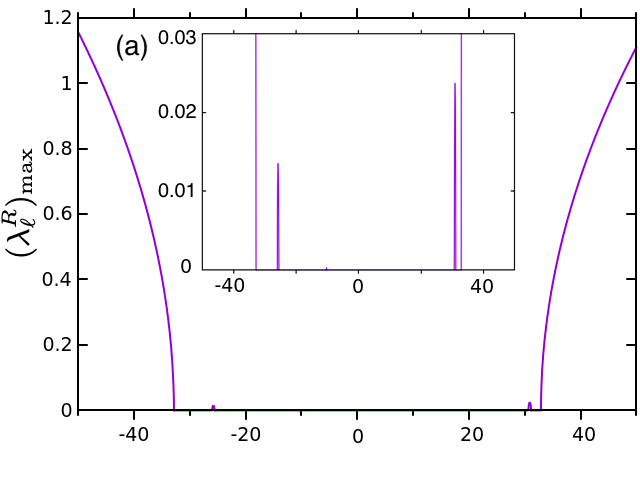}\vspace{-0.3cm}
     \includegraphics[scale=0.48]{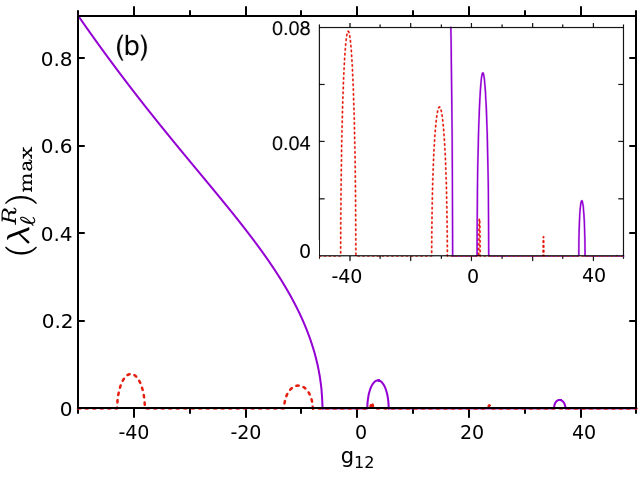}
     \vspace{-0.2cm}
\caption{(Color online) Floquet spectra, given by $(\lambda_{\ell}^R)_{\rm max}$, 
from \eqref{floquet_exp}, are shown as functions of the inter-species interaction
$g_{12}$, with the intra-species and Rabi coupling parameters fixed at 
($\Omega$, $g$)$=$ (0.1, 20) (a) and (0.9, 40) (b). 
The $\ell=$1 and 2 unstable modes are, respectively, represented by solid-violet
and dashed-orange lines. In the insets,$(\lambda_{\ell}^R)_{\rm max}$ is re-scaled
to improve visibitily of lower peak instability regions. 
Within defined units, all quantities are presented as dimensionless.
}
\label{fig05}
\end{figure}

Figure~\ref{fig04} displays two panels with Floquet stability diagrams, parametrized
by the intra- and inter-species interactions, obtained from 
 the oscillating functions $f_{j}$ for a complete period $T$. 
 In panel (a), we choose a small value for $\Omega=0.1$, that provides 
 good agreement with the stationary results presented in panel (a) of 
 Fig.~\ref{fig03}. As shown in panel (b), the agreement is no more maintained
when considering a large value $\Omega=0.9$ for the Rabi constant, 
 as compared with panel (b) of  Fig.~\ref{fig03}.
 In these diagrams, the dominant unstable angular modes $\ell=1,2,3$ are 
 indicated inside the panels. Besides the similarity between the diagrams (a)
 of Figs.~\ref{fig03} and \ref{fig04}, for $\Omega=0.1$, the corresponding diagrams
 are already useful to verify the effect of more detailed stability analysis.
As noticed, some stable regions verified with the BdG approach are no more 
confirmed when using the Floquet method, such as the regions with $g<0$, near 
$g_{12}=0$. Even for  $g>0$, in the dominantly stable regions pointed out by 
the BdG approach,  we can already verify instabilities detected by the Floquet method.
Particularly, the border of the regions can no longer be kept when we carry out 
a more accurate stability study. 
These results, together with the following ones that we are going to discuss,
led to the conclusion that, as soon as the Rabi coupling is turned on, 
the Floquet method is more sensible to system instabilities
as one varies the inter- and intra-species interactions, being more
accurate in studying the stability of a system than the conventional 
stationary BdG approach.

In the two panels of Fig.~\ref{fig05}, we select two sets of parameters from 
the Floquet diagrams shown in Fig.~\ref{fig04}, for separate representation of 
the corresponding spectra, given by the maximum of the real part of
the Floquet exponent, $(\lambda_\ell^R)_{\rm max}$. 
The two panels are representing the respective spectrum in terms of the
inter-species interaction $g_{12}$ by considering fixed values of  
intra-species interactions and Rabi frequencies, with ($\Omega$, $g$)= (0.1, 20) in 
panel (a) and (0.9, 40) in panel (b).
Figure~\ref{fig05}, with the spectrum of unstable modes, is also indicating 
the meaning of the very faint lines appearing in the two diagrams 
of Fig.~\ref{fig04}. In these two cases, we have only unstable modes with 
$\ell=$ 1 (solid-violet lines) and 2 (dashed-orange lines), as indicated.
The respective insets in both panels are displayed to enhance the visibility of the
lower peaks observed in the larger panels.
The Floquet spectrum is also able to predict the existence of resonant 
conditions that can happen according to the chosen parameters. The conditions
for that will be discussed in the next section \ref{sec4_sub}, where 
a comparison is provided with a semi-analytic model when $|\Delta g|/(8\pi)\ll\Omega$.

\subsection{Resonance Conditions}\label{sec4_sub}
It is possible to figure out the excitation mechanism responsible for the observed 
Floquet unstable spectrum by analyzing the resonance conditions, as discussed 
in Ref.~\cite{2019Chen}. Our approach mainly differs from this reference when 
considering the free-particle spectrum in the formalism, as in our case the 
full kinetic energy term is provided by the squared angular momentum operator. 
Therefore, the continuum $\epsilon_k$ must be replaced by the corresponding 
discrete angular spectrum $\epsilon_\ell$. 
So, with the assumption that $|\Delta g|\ll 8\pi\Omega$, by using a first-order 
approximation in $\Delta g$ with slight corrections in the solutions 
in Eq.~\eqref{eq03}, valid in the regime $g_{12}=g$, we are able to get 
linearized equations (See Appendix~\ref{app_resonance_v2}), with 
two natural frequencies as obtained from \eqref{C6}. In the limit 
$|\Delta g|=|g_{12}-g|\ll 8\pi\epsilon_\ell$, they are
{\small\begin{subequations}\begin{eqnarray}
&&\omega_{d, \ell}\approx
\sqrt{\epsilon_\ell\left(\epsilon_{\ell}+\frac{g}{2\pi}\right)}+
\frac{\sqrt{\epsilon_\ell}\Delta g}{16\pi\sqrt{
{\epsilon_{\ell}+\frac{g}{2\pi}}}}\label{res-freqa}, \\
&&\omega_{s, \ell}\approx
\epsilon_\ell+\frac{\Delta g}{16\pi}
\label{res-freqb}.
\end{eqnarray}\end{subequations}
}
\begin{figure}[h]
    \centering    \includegraphics[scale=0.5]{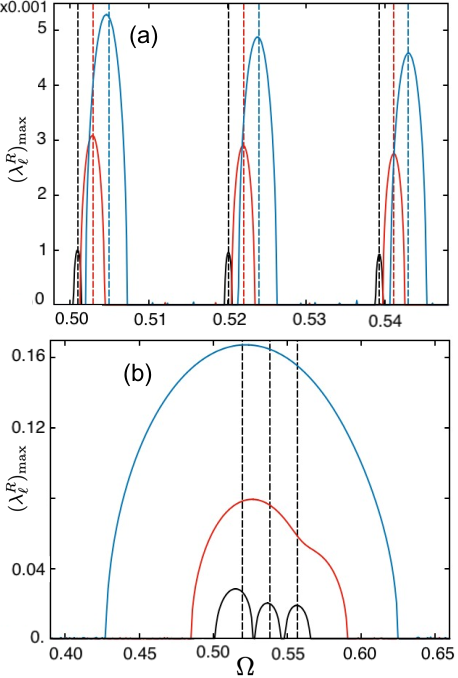}
    \vspace{-0.2cm}
\caption{(Color online)
 Unstable Floquet spectra [$(\lambda_\ell^R)_{\rm max}$], for
$\ell=1$ with different $\Delta g$, are shown as functions of the coupling $\Omega$. 
In (a), the results are for 
$\Delta g=$ 0.1 (the three black curves with maxima near 0.001), 
0.3 (the three red curves with maxima near 0.003), and
0.5 (the three blue curves with maxima higher than 0.004).
In (b), we consider higher $\Delta g$ values, with
$\Delta g=$ 2.0 (the three black curves with maxima below 0.03),
4.0 (the single red curve with a maximum near 0.08),  and 
7.0 (the larger single blue curve with a maximum near 0.17). 
The dashed vertical lines, obtained from (\ref{om123}), 
are pointing out the corresponding predictions, which are very 
close to the respective maxima, in (a). However, in (b), the 
close agreement happens only for $\Delta g=$ 2.0, as shown.
Within defined units, all quantities are presented as dimensionless.}
\label{fig06}
\end{figure} 
Parametric resonances are achieved when an external potential go with about 
twice the natural frequencies of the system \cite{1960Landau}. Once the $\ell$-mode 
excitations evolve in time with $\cos(4\Omega_{\ell} t)$ and 
$\sin(4\Omega_{\ell} t)$~\cite{2019Chen}, three critical couplings 
$\Omega_{\ell}$ emerge, which can be tuned to trigger the resonances
\begin{eqnarray}
{\Omega}_{\ell}^{(1)}=\frac{\omega_{d, \ell}}{2},\;\; 
{\Omega}_{\ell}^{(2)}=\frac{\omega_{s, \ell}}{2},\;\;
{\Omega}_{\ell}^{(3)}=\frac{\omega_{d, \ell}+\omega_{s, \ell}}{4},\label{om123}
\end{eqnarray}
which are usually associated with density-density, spin-spin, and density-spin 
resonances, respectively \cite{2019Chen}. For the limit $g=0$, one can clearly 
see that the three resonant peaks are going to merge in just one peak,
$(1/2)\left({\epsilon_\ell}+\frac{g_{12}}{16\pi}\right)$.
These resonant positions can be observed in the Floquet spectrum in the regime 
of $g_{12}\approx g$. When $g_{12}$ becomes higher, 
the three peaks continuously merge into only one peak. In Fig.~\ref{fig06}, we 
compare the Floquet spectrum with the approximations given by \eqref{om123} for 
the resonance couplings. As seen in panel (a), for $\Delta g=$0.1, 0.3, and 0.5, the 
predicted values match exactly with the resonance peaks. In panel (b), the
three-peak predictions are shown only for $\Delta g=$2.0, which are close to the 
exact numerical solutions. In the other two cases, with $\Delta g = $ 4 and 7, 
as the predictions are no more valid, we include only the numerical exact solutions 
presenting the corresponding single maxima.

\section{Atom-population dynamics}\label{sec5}
The dynamics of the atom-population exchange for the system was done by
full numeric calculations of the coupled GP Eqs.~(\ref{eq01}),
carrying out the spectral method introduced in Ref.~\cite{2021Andriati}. 
Within our numerical computation, the dynamics is performed with time 
steps $\Delta t=10^{-5}$, having spatial grids, in $\theta$ and 
$\phi$ directions, with range sizes of $256\times256$ and respective
step sizes given by $\Delta\theta=\pi/256\approx0.013$ and
$\Delta\phi=2\pi/256\approx0.025$.
The GP equations are solved by starting with homogeneous solutions in which
each species has half of the total population, i.e, 
$\psi_{1}=\psi_{2}=1/\sqrt{8\pi}$, with a $5\%$ random noise added to each
point in the mesh grid. 
In this numerical approach, we are able to verify how long the 
homogeneous periodic solutions (\ref{eq04}) solved by using the RK4 method
provide a good model to describe the evolution of the populations. 
The stability behavior of the homogeneous miscible initial states is observed by 
displaying their overlap evolution, population dynamics, and density pattern
when unstable modes occur.
\begin{figure}[h]
    \centering    \includegraphics[scale=0.4]{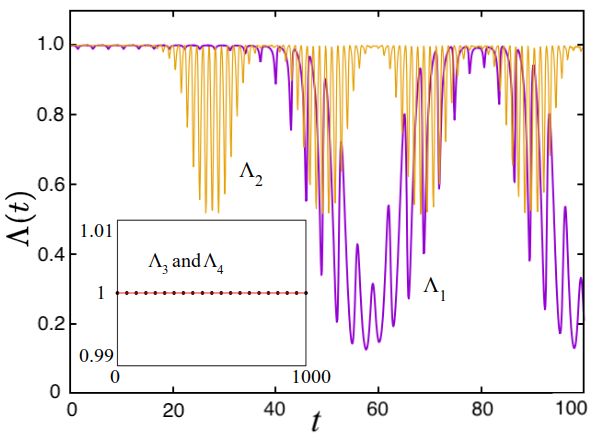}
    \vspace{-0.2cm}
\caption{(Color online) Time-evolution of density overlaps, $\Lambda(t)$, from (\ref{overlap_eq}), for the coupling and 
set interactions ($\Omega, g, g_{12}$) given by:
$\Lambda_1$ (0.50, 1, 8), $\Lambda_2$ (0.94, 40, -10),
$\Lambda_3$ (0.10, 1, 10), and $\Lambda_4$ (0.99, 1, 25).
$\Lambda_1$ and $\Lambda_2$ (main panel) are identified by 
solid-violet and solid-orange lines. The stability of $\Lambda_{3, 4}$ is
shown in the inset. Within defined units, all quantities are presented as dimensionless.}
\label{fig07}
\end{figure}
\begin{figure}[h]
       \includegraphics[scale=0.48]{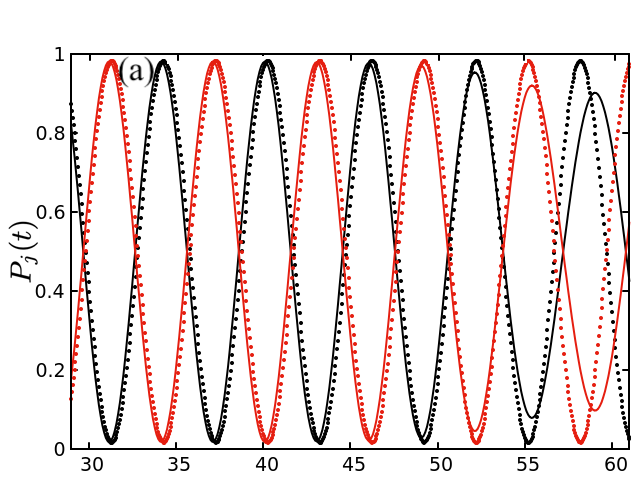}
\vspace{-0.2cm}       \includegraphics[scale=0.48]{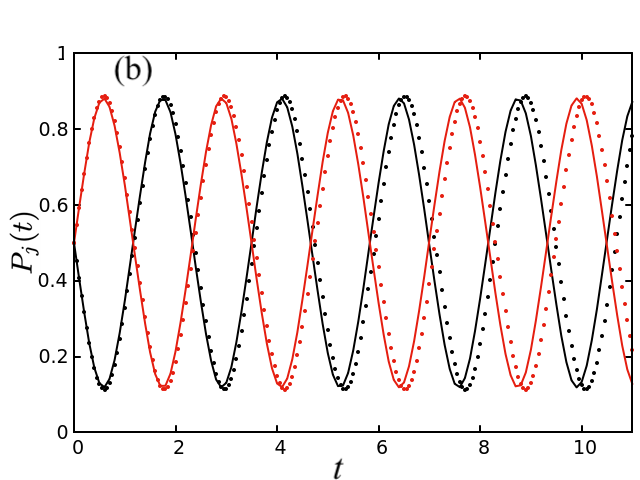}
\caption{(Color online) The population time-evolution $P_{j}(t)$, given by 
(\ref{population_eq}), is shown for the Rabi and interaction parameters 
($\Omega$, $g$, $g_{12}$) = (0.50, 1, 8) (a) and (0.94, 40, -10) (b), 
respectively. Solid lines stand for the full GP computation of \eqref{eq01}, 
which takes into account the spatial-time-dependent wave functions, with the 
dotted ones for the homogeneous only time-dependent solutions \eqref{eq04}. 
The colors, black and red, are 
identifying, respectively, species 1 and 2. The amplitudes agree with \eqref{amplitude}, 
being ${\cal A}=0.9657$, for $\Delta g= 7$; and ${\cal A}=0.7737$, for $\Delta g = 50$. 
Within defined units, all quantities are presented as dimensionless.
}\label{fig08}
\end{figure}
\begin{figure}[h]
    \centering
       \includegraphics[scale=0.48]{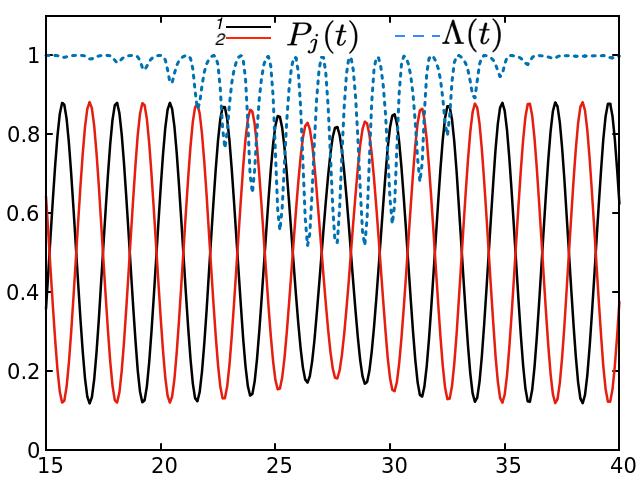}
  \includegraphics[scale=0.48]{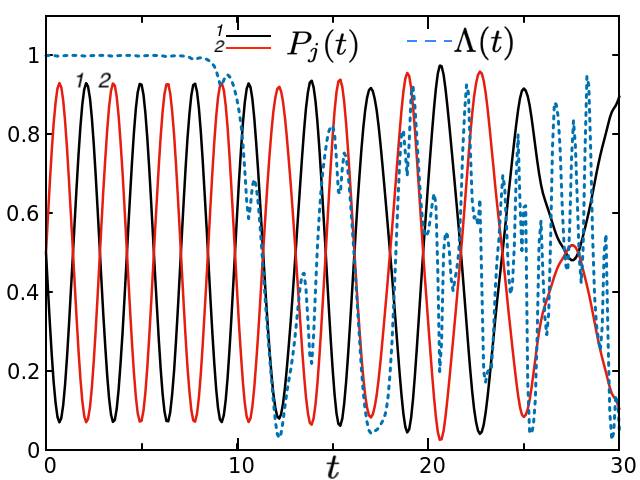}
     \vspace{-0.2cm}
\caption{(Color online) Time-evolution of populations, $P_{j}(t)$ 
(solid lines) and overlaps, $\Lambda(t)$ (dashed-blue lines), as given 
by (\ref{population_eq}) and (\ref{overlap_eq}),respectively. Species 1 and 
2 populations are given by solid lines (black and red, respectively), as indicated.  
The Rabi and interaction parameters are 
($\Omega$, $g$, $g_{12}$) = (0.94, 40, -10) in (a); and
(0.90, -10, 20), in (b). Within defined units, all quantities are presented as dimensionless.}
\label{fig09}
\end{figure}

To estimate the miscibility of the system, the overlap 
of densities are verified by the parameter $\Lambda$, 
defined by
{\small
\begin{equation}
    \Lambda(t)=\frac{{\left[ \int d\phi \sin\theta d\theta 
    {|\psi_1|}^{2}{|\psi_2|}^{2}\right]}^{2}}
    {\left[\int d\phi \sin\theta d\theta 
    {|\psi_1|}^{4} \right]\left[\int d\phi \sin\theta d\theta 
    {|\psi_2|}^{4} \right]}.
    \label{overlap_eq}
\end{equation}
}When $\Lambda= 1$, the species are miscible, while $\Lambda<1$ stands for immiscible 
condensates. Once the initial overlap decreases, it means that the initial miscible 
setup is no longer stable. In Fig.~\ref{fig07}, we show the overlap dynamics regarding 
four different set of parameters, where the set of parameters including intra- and 
inter-species interaction, and Rabi coupling constant are given by ($\Omega$, $g$, 
$g_{12}$)=(0.50, 1, 8), (0.94, 40, -10),
(0.10, 1, 10), and (0.99, 1, 25), for which the stability predictions can be localized in 
Figs.~\ref{fig13}{(b)}, \ref{fig15} , \ref{fig04}{(a)} and \ref{fig13}{(c)}, respectively. 
A complementary analysis can be made by observing the population 
dynamics of the previous cases, with the population $P_{j}$ of each species
given by 
{\small
\begin{equation}
    P_{j}(t)=\frac{\int d\phi \sin\theta d\theta \ 
    {|\psi_j|}^{2}}{\int  d\phi \sin\theta d\theta \ 
    \left[|\psi_1|^{2}+|\psi_2|^{2}\right] } = \frac{N_j(t)}{N}.
    \label{population_eq}
\end{equation}}

Figure~\ref{fig08} shows how the population oscillation is affected when 
the system becomes unstable. The population behavior is closely related to the overlap 
since both properties are changed when the miscible homogeneous initial ansatz 
(\ref{eq04}) are no longer the true solutions of the system. It is important to note 
that the overlap dynamics for unstable cases are driven by two different frequencies. 
The slow frequency is a periodic behavior of miscibility, which was first observed 
in our previous work \cite{2021Andriati}, and it happens only for specific choices 
for interaction parameters. Moreover, in this work, we observe a second frequency in 
the overlap dynamics, which is faster than the first one, and is driven by 
the population dynamics frequency. In Fig.~\ref{fig09}, we present 
two different sets of parameters where both frequencies are actually leading 
the overlap behavior. In panel \ref{fig09}{(a)}, which refers to 
($\Omega$, $g$, $g_{12}$) = (0.94, 40, -10), we clearly see that the faster 
kind of overlap oscillation has the same frequency as the population dynamics. 
Another set of parameters is depicted in panel \ref{fig09}{(b)},
with
($\Omega$, $g$, $g_{12}$) = (0.90, -10, 20), for which we have the stability 
prediction in Fig.~\ref{fig04}{(b)}. Here, observing some periodic 
behavior with two distinguished frequencies is not as direct as in the previous case. 
This is an example where slower kind of modulation is not periodic, as also being observed
\cite{2021Andriati}.
In this way, the periodic oscillation caused by the Rabi coupling can be expected 
for all choices of parameters, but the same statement is not true for slower 
kind of modulation. 
\begin{figure}[!htb]
    \centering
     \includegraphics[scale=0.5]{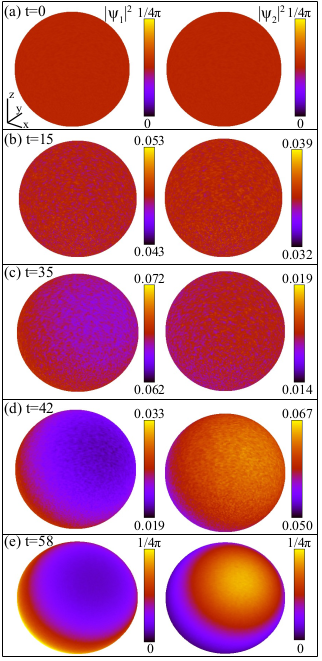}
\vspace{-0.2cm}
\caption{(Color online)
Dynamics of densities ${|\psi_{1}|}^{2}$ (left column) and ${|\psi_{2}|}^{2}$ (right column), for the species $j=1$ and 2,
considering the Rabi and interactions given by ($\Omega$, $g$, $g_{12}$) = (0.50, 1, 8).
The time snapshots $t$ are indicated inside respective pair of panels, with density variations, according to \eqref{eq06} and \eqref{amplitude}, being within the interval $[0.0014\le|\psi_j|^2\le 0.0782$. At any $t$, 
$\sum_j|\psi_j|^2={1}/{(4\pi)}$, with
identical $|\psi_j|^2=1/(8\pi)$ at $t=0$ and at each half period $T/2$. 
Within defined units, all quantities are presented as dimensionless. }
\label{fig10}
\end{figure}
\begin{figure*}[!htb]
    \centering
    \includegraphics[scale=0.47]{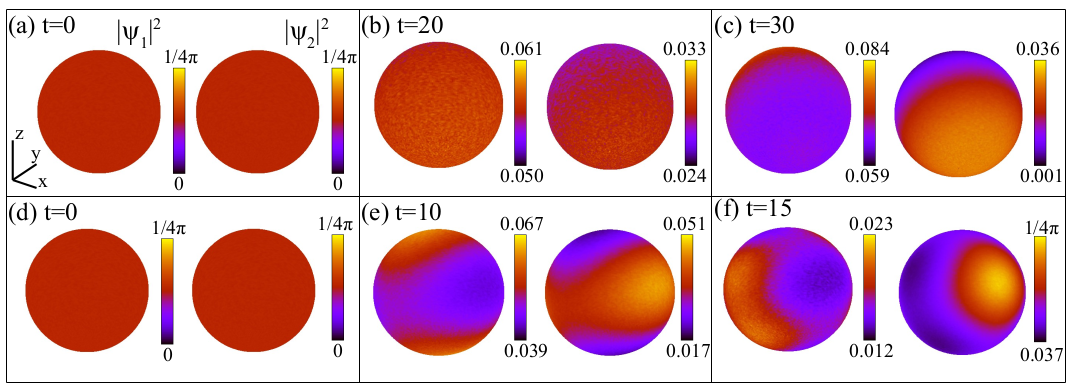}
\vspace{-0.2cm}
\caption{(Color online) Dynamics of densities ${|\psi_{j}|}^{2}$ for the parameters 
($\Omega$, $g$, $g_{12}$) = (0.94, 40, -10)
[upper panels,  
with $t$ = 0 (a), 20 (b), and 30 (c)], and 
(0.9, -10, 20) 
[lower panels,  
with $t$ = 0 (d), 10 (e), and 15 (f)].   
 The range of densities in the upper and lower panels are, respectively,
$[0.01388\le|\psi_j|^2\le 0.06569]$ and $[0.00529\le|\psi_j|^2\le 0.07428]$. For each pair,
the densities $|\psi_j|^2$ are given as indicated 
inside panel (a).
Within defined units, all quantities are presented as dimensionless.
}
\label{fig11}
\end{figure*}
\begin{figure}[!htb]
    \centering
    \includegraphics[scale=0.48]{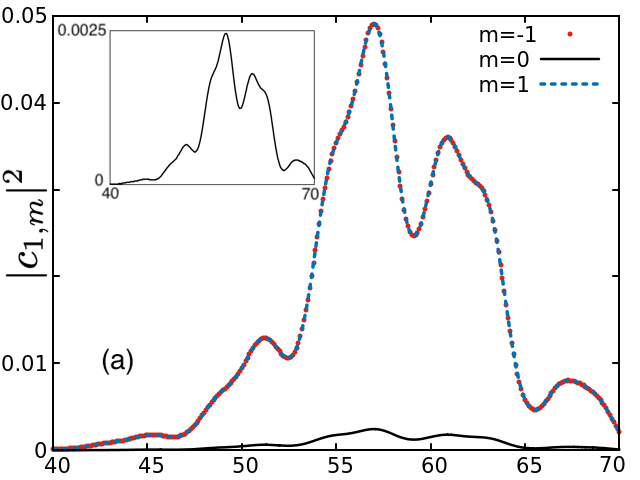}
        \includegraphics[scale=0.48]{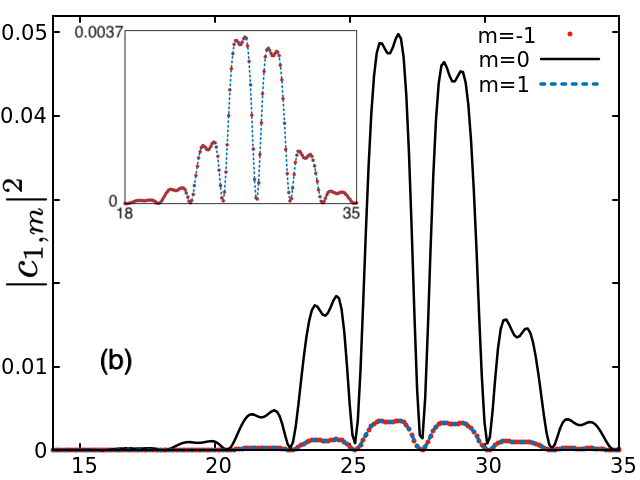}
            \includegraphics[scale=0.48]{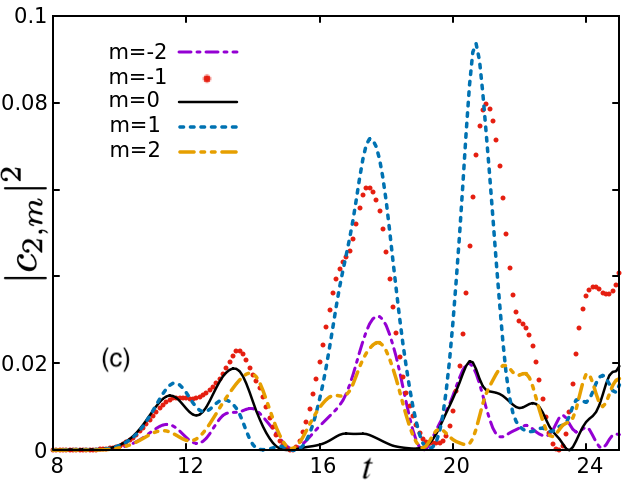}
            \vspace{-0.2cm}
\caption{(Color online) Time evolution of couplings ${|c_{\ell, m}(t)|}^{2}$ for the species 1 [see \eqref{coup_eq}],
for three unstable different set 
of parameters ($\Omega$, $g$, $g_{12}$) = (0.50, 1, 8), (0.94, 40, -10), and (0.90, -10, 20), which are depicted in (a), (b), and (c), respectively. The insets in (a) and (b) stand 
for the lower lines shown in the main panels.
Within defined units, all quantities are presented as dimensionless. 
}
\label{fig12}
\end{figure}
The dynamics due to an unstable mode driving the behavior of the system 
can be more clearly seen in Fig.~\ref{fig10}, where we 
are displaying the time-evolution of the densities,
with the parameters ($\Omega$, $g$, $g_{12}$)=(0.50, 1, 8), for which 
unstable behavior is being predicted to happen in the angular mode $\ell=1$
{[See panel (b) of Fig.~\ref{fig13} in the Appendix~\ref{app_comparison}]}. 
By observing the 
density dynamics, we are able to see that, at some time, a density pattern 
can emerge in both species, which soon evolves into an immiscible setup,  
where the condensates of each species reduce to localized small clouds, 
and therefore, Faraday waves become difficult to be observed. 

A similar calculation is performed in Fig.~\ref{fig11}, where we show how 
the densities evolve for the unstable sets of parameters ($\Omega$, $g$, $g_{12}$)= 
(0.94, 40, -10) and (0.9, -10, 20). The first row of the figure shows how the Faraday patterns 
emerge, with the two species going to an immiscible-phase configuration of localized small pieces.
In the second row, it is noticed that both condensates are soon breaking into two pieces, 
where also emerge Faraday patterns. The stability of both sets can be checked by the 
Floquet spectrum in panels \ref{fig04}{(a)} and Fig. \ref{fig15} 
in the Appendix \ref{app_comparison},  respectively. Which predicts that both cases 
are unstable, and driven by the modes $\ell=1$ and $\ell=2$, respectively. The dynamics 
simulations confirm these predictions, and in each case, the condensates are likely to 
become small localized clouds and break into two pieces, respectively. Therefore, 
the Floquet spectrum correctly 
predicts the stability behavior observed in the dynamics. 

Our present analysis is extended to the Appendix \ref{app_comparison}, in which 
the BdG and Floquet stability predictions are being compared with full dynamical
results. 
We figure out that, as soon as $\Omega>0$, the Floquet method offers a more reliable 
stability profile for homogeneous time-periodic states, with the BdG spectrum 
returning similar results only low coupling constants $\Omega\ll 1$. 
Following these analyses, we can summarize by illustrating the density dynamics 
simulations of some unstable cases displayed in Figs.~\ref{fig10} 
and \ref{fig11}. These results show us that, once 
an unstable angular mode $\ell$ takes over the dynamics of the system,
the condensates will break into the corresponding  number $\ell$ of localized immiscible
pieces \cite{2021Andriati}. Here, we are pointing out that these 
angular modes are also able to provoke the emergence of Faraday waves, by
tuning the Rabi coupling to the
natural resonance frequencies of elementary excitations.
Nevertheless, phase separations are expected to happen on a higher scale 
of densities, with possible 
localized small condensate clouds breaking, within
a condition that Faraday-wave effects are not likely to be seen.

In Fig.~\ref{fig12}, we quantify the effect of the unstable modes 
on the dynamics by the square modulus of the coupling $|c_{\ell,m}|^2$,
where the coefficients $c_{\ell, m}$ are given by
\begin{equation}
    c_{\ell, m}(t)=\int 
    d\phi \sin\theta d\theta \ {Y}_{\ell, m}^{*}(\theta,\phi)
    \psi_{j}(\theta,\phi,t).
    \label{coup_eq}
\end{equation}
We calculate the coupling for the species 1 wave function coupled with 
the spherical harmonics $Y_{\ell, m}(\theta,\phi)$. 
We observe three unstable cases, for which we find out that once an angular 
mode $\ell$ is unstable, the amplitudes of the couplings regarding 
each degenerate mode $m=-\ell,\cdots,\ell$ are arbitrary. Note that 
in the third case, depicted on the panel \ref{fig12}{ (c)}, 
we show only the coupling with degenerate modes associated with 
$\ell=2$, since they are the dominant ones, and the early modes to drive 
the dynamics. Modes regarding $\ell=1$ also can be important for 
longer times. For cases with $\ell=1$, the coupling with the modes 
$m=-1,0,1$ has the same behavior, but the mode $m=0$ 
has a different amplitude of the modes $m=\pm1$. When $\ell>1$, 
this symmetric behavior between the modes $m$ is no longer observed.

\section{Conclusions}\label{sec6}
The dynamics and stability of homogeneous binary BEC mixtures trapped on a
spherical bubble are investigated by considering atom-number oscillations achieved by Rabi 
coupling.
Exact analytical solutions are developed for population dynamics, followed by 
stability analyses considering BdG and Floquet methods, which are compared with
the corresponding full numerical solutions.
In the stability analyses, we first examined the role of Rabi coupling on 
stationary solutions by applying the BdG method. This was followed by a more detailed analysis
of the associated dynamics by using the time-dependent Floquet method.

As concerning to the methods applied for studying the stability analysis, our approach is similar 
to~\cite{2019Chen}. However, both 2D confining systems have quite different characteristics, 
from the physical and numerical point of view. 
Within an infinite surface plane, the authors of~\cite{2019Chen} had a continuum kinetic 
spectrum to study the production of Faraday patterns by periodic modulations of the
effective interaction, whereas in the present case, with fixed radius leading to a discrete 
kinetic energy spectrum,  
the parametric resonances are achieved by modulating the Rabi frequency
in a 2D spherical system within periodic-boundary conditions.
As observed, the Rabi oscillations are able to drive the system to different stability 
profiles, once an effective time-oscillating interaction energy is performed. 
In this kind of 2D spherical topology, discrete unstable orbital angular modes can 
rise and lead the BEC mixture to an immiscible phase separation, in which the condensate 
can break into a corresponding discrete number of localized clouds. 
Since there is an effective interaction modulation, it is relevant to note that 
the unstable degenerate azimuthal angular modes can give rise to Faraday waves, which 
coexist with the separate phase. As shown for some range of parameters, the system can 
enter a periodic regime, where the miscibility of the species can vary in time, dynamically. 

As perspectives for further investigations, of particular relevance is to
consider a more general 3D study regarding spherical topology, 
in which the radial skin becomes a parameter in the theory. 
As in this case only discrete modes are allowed, a phase separation where 
the coupled condensate breaks into localized fixed-number of clouds presents 
a density order much higher than the Faraday wave patterns. Eventually, the 
Faraday wave phenomenon can be hidden within this process in which the 
breakdown of the clouds turns out to be too much faster. 
As other possible extensions, within the same spherical geometry context,
one could study the stability of dipolar-coupled systems or how nonlinear 
quantum fluctuations could affect the outcome of this work.

Finally, besides not being reported up to now dual-species BEC 
mixtures in the ultracold bubbles experimentally achieved in microgravity conditions, 
to this aim a possible way is to exploit atomic mixtures with their tunable interaction
(as discussed in~\cite{2023Lundblad}). 
Also noticeable in this regard is the fact that the original trap proposal for matter-wave 
bubbles~\cite{2001Zobay} was based on driving adiabatic potentials with 
Rabi-coupled hyperfine-states, pointing out the impact
of the present and related theoretical analyses.
Our work can provide some insights on how it is 
possible to trigger parametric resonances, or even avoid them when dealing 
with cold-atom state mixtures. Also, in our approach, noticeable is the fact that 
the only parameters needed to drive the occurrence of Faraday wave resonances are 
the $s-$wave nonlinear interactions and the Rabi coupling. 

\begin{acknowledgements}
LB thanks Prof. Hiroki Saito for useful discussions. 
The authors acknowledge the Brazilian agencies Funda\c{c}\~ao de Amparo \`a 
Pesquisa do Estado de S\~ao Paulo (FAPESP) [Contracts 2017/05660-0 (LT), 
2016/17612-7 (AG)], Conselho Nacional de Desenvolvimento Cient\'\i fico e 
Tecnol\'ogico [Procs. 304469-2019-0 (LT) and 306920/2018-2 (AG)] and 
Coordena\c{c}\~ao de Aperfei\c{c}oamento de Pessoal de N\'\i vel Superior 
[Proc. 88887.374855/2019-00 (LB)].
\end{acknowledgements}

\appendix
{\section{3D to spherical 2D dimensional reduction and adimensionalization}\label{2D_red+adim}
The formalism reduction from 3D to the spherical 2D, for coupled condensates
trapped in a fixed-radius bubble, is performed in this section, starting from
the full-dimensional space-time variables ($\tilde r,\tilde t$) and parameters.
Once the formalism is in 2D format, we show how the adimensionalization leads 
to \eqref{eq01}. The wave function for the two species $j$,  normalized to 
$N_j$, are given by $\Phi_j\equiv\Phi_{j}(\tilde{r},\theta,\phi;\tilde{t})$,
such that $N=N_1+N_2$. With both species having the same mass $M$, coupled by 
a Rabi oscillating frequency $\Omega_R$, confined radially by a common 
symmetric potential $V(\tilde{r})$ ($=0$ for $R-\delta R/2<\tilde{r}<R+
\delta R/2$; and $\infty$ otherwise), we obtain the following time-dependent
coupled formalism:
{\begin{eqnarray}
{\rm i}\hbar\frac{\partial}{\partial\tilde{t}}\Phi_{j=1,2}
&\!=\!&\left[-\frac{\hbar^2}{2M}\nabla^2 + V(\tilde{r})\right]\Phi_{j}
+(-1)^{j}{\mathrm i}\hbar\Omega_R\Phi_{3-j}
\nonumber\\
&+&\sum_{i=1,2}
\overline{g}_{ji}
{\left| \Phi_{i}\right|}^{2}
\Phi_{j}, 
\label{apA01}
\end{eqnarray}
}where $0\leq \tilde{r}<\infty$, $\theta\in[0,\pi]$ and $\phi\in[0,2\pi]$, with 
$\overline{g}_{ij}\equiv4\pi\hbar^2{a}_{ij}N/M$, by assuming the
total wave function is normalized to one.
 With the system confined at the surface of a large bubble having fixed radius $R$, the
 radial part of the formalism can be solved by using a common ansatz ${\cal R}(\tilde{r})$ for both species, which must vanish outside
 a skin with thickness $\delta R\ll R$, for the 3D spherical shell. So, with the full dynamics given by the angular part and the time $\tilde{t}$, we assume a $\delta-$like Gaussian shaped form for 
 the radial part of $\Phi_j$, in \eqref{apA01}, such that
 $\Phi_j\equiv{\cal R}(\tilde{r})\Psi_{j}(\theta,\phi;\tilde{t})$, with
${\cal R}(\tilde{r})\equiv \displaystyle\frac{1}{\sqrt{\sigma\sqrt{\pi}}R}\exp\left[\frac{-(\tilde{r}-R)^2}{2\sigma^2}\right]$ normalized as $\int_0^\infty d\tilde{r} \tilde{r}^2 [{\cal R}(\tilde{r})]^2=1$, where the Gaussian width $\sigma$ can be directly identified with the thickness $\delta R$.
Once integrated the radial part, and by identifying $\sigma=\delta R$, such that $g_{ij}\equiv \frac{\sqrt{8 \pi}a_{ij}N}{\delta R}$, we obtain
{\small\begin{eqnarray}
\hspace{-0.7cm}{\rm i}\hbar\frac{\partial\Psi_{j}}{\partial \tilde{t}}\!\!&\!\!=\!\!&\frac{\hbar^2}{MR^2}\!\!\left[\frac{{\bf L}^2}{2\hbar^2}+\!\!
\sum_{i=1,2}\!\!\!g_{ji}
{\left| \Psi_{i}\right|}^{2}\right]\!\!\Psi_{j}
+(-1)^{j}{\mathrm i}\hbar\Omega_R\Psi_{3-j},
\label{apA02}
\end{eqnarray}
}where ${\bf L}$ is the angular momentum operator.
Next, for the adimensionalization, we should first notice that $\Psi$ has
only angular dependence. So,  
we can simply assume $R$ as
our length unit, such that $\hbar^2/(MR^2)$, $MR^2/\hbar$ and 
$\hbar/(MR^2)$ will be, respectively, the energy, time and 
frequency units. Within these units, $\delta R$ will be understood as an infinitesimal
$\delta$ times $R$, such that $g_{ij}$ will be dimensionless, and 
the Rabi oscillating 
parameter is given by 
$\Omega_R=\Omega [\hbar/(MR^2)]$. Finally, by factoring the energy unit
in \eqref{apA02},
we end up with \eqref{eq01}.
}

\section{Binary density oscillations}\label{app_density}
This Appendix is concerned with the exact time-dependent behavior of the 
dimensionless coupled formalism \eqref{eq01},
by assuming~\eqref{eq04}, 
where $f_j$ and $\gamma_j$ are to be found considering the 
interactions $(g, g_{12})$ and Rabi constant $\Omega$. Here,  
$f_j$ are assumed real with time-independent $\gamma_j$, 
considering any possible time dependence of the phases 
provided by redefinitions of  
the wave-function phases in \eqref{eq04}, with 
$\psi_j=f_j e^{-{\rm i}[\gamma_j t+\beta_j(t)]}$
(To simplify the notation, we start using upper-dot notation 
or the suffix $t$ for the time derivatives).
So, from \eqref{eq04} in \eqref{eq01}, and with the complex phase written as
$\Delta_{\gamma\beta}\equiv(\gamma_2-\gamma_1) t+[\beta_2(t)-\beta_1(t)]$, and
also with the redefinition $\bar\gamma_j(t)\equiv\gamma_j+\dot\beta_j(t)$, we obtain
{\small
\begin{eqnarray}
{\rm i}\partial_t f_j\!\!=\!\![G_{j,3-j}-\bar\gamma_j]f_j
+(-1)^{j}{\rm i}\Omega f_{3-j}
e^{(-)^j{\rm i}{\Delta_{\gamma\beta}}},
\label{gp2}
\end{eqnarray}
}where 
$G_{j,3-j}\equiv g |f_j(t)|^2+g_{12}|f_{3-j}(t)|^2$.
After separating the real and imaginary parts and rearranging
the terms, we obtain the harmonic-like oscillator equation, with 
density-dependent frequency, given by
{\small\begin{eqnarray}
\left[\partial_{tt}+
\Omega^2\!-\!(\bar\gamma_j-G_{1,2})
(\bar\gamma_j- G_{2,1})\right]f_j\!=\!0.
\label{W}
\end{eqnarray}
}The initial condition $|f_j(0)|^2=1/(8\pi)$ with the frequency  
reduced to $\Omega^2$ leads to 
{\small
\begin{equation}
\bar\gamma_j(0)=\gamma_0=\frac{g+g_{12}}{8\pi},\;\;{\rm with}\;\;
\dot\beta_j(0)=0
\label{gamma},
\end{equation}
}implying that $\bar\gamma_j$ has a constant part $\gamma_0$ and a time-dependent
part $\dot\beta_j$. By replacing $\bar\gamma_j$ in \eqref{W}, we have 
{\small\begin{eqnarray}
&&\partial_{tt} f_{j}+W_j^2(t)f_j
=0, \;\;{\rm with}\nonumber\\
&&W_j^2(t)\equiv
\left[\Omega^2+\left(\frac{\Delta g}{2}\right)^2\left({|f_2|^2-|f_1|^2}\right)^2-\left(\dot\beta_j\right)^2\right].
\label{w2t}\end{eqnarray}
}In the limiting cases in which $g=g_{12}$, the solutions are already known 
being sinusoidal with the oscillation given by $\Omega$.  However, when 
$g\ne g_{12}$, the frequency depends on the square of the differences between 
the two condensates, with the interval of oscillations for 
$(|f_2|^2-|f_1|^2)$ being reduced, 
It goes from 0 to $\left|{\cal A}\right|$,
with each density oscillating from $\frac{1}{4\pi}-|{\cal A}|$ to $|{\cal A}|$, 
constrained by normalization and initial conditions. 
It is convenient to solve \eqref{w2t} for the
density difference (or atom-number difference), $\nu(t)\equiv\frac{N_2(t)-N_1(t)}{N}$
$={4\pi}{(|f_2|^2-|f_1|^2)}$. So, we follow from \eqref{eq01}, with an explicit 
derivation of the equation for $\nu(t)$, starting with its first derivative 
{\small\begin{eqnarray}
\partial_{t}\nu= 8\pi\Omega\left[\psi_1^*\psi_2 + \psi_2^*\psi_1\right],
\label{nu1}\end{eqnarray}
}followed by the 2nd derivative
\begin{equation}
\partial_{tt}\nu=8\pi\Omega\left(\psi_2{\partial_t \psi_1^*} + \psi_1^*{\partial_t \psi_2}\right)
+\text{c.c.}\label{E6}
\end{equation}
After some straight manipulations, we obtain
\begin{equation}
\partial_{tt}\nu+4\Omega^2\nu=2\Omega {\mathrm i}(\Delta g)
\nu(\psi_1^*\psi_2-\psi_1\psi_2^*), \label{E7}
\end{equation} 
in which the right-hand-side can be solved through the corresponding derivative,
\begin{eqnarray}
\partial_t\left(\psi_1^*\psi_2-\psi_2^*\psi_1\right)&=&{\mathrm i}
\frac{\nu}{4\pi}(\Delta g)
\left(\psi_1^*\psi_2+\psi_1\psi_2^*\right)\nonumber\\ 
&=&{\mathrm i}\frac{\Delta g}{16\pi\Omega}\partial_t(\nu^2).
\label{E10}
\end{eqnarray}
By integrating both sides from $0$ to $t$, and using the initial conditions at $t=0$, with
$\nu(0)=0$ and the $\psi_1(0)=\psi_2(0)$, 
{\small
\begin{align}
\left(    \psi_1^*\psi_2-\psi_2^*\psi_1\right)= 
{\mathrm i}\frac{\Delta g}{16\pi\Omega}(\nu^2).
    \label{E10c}
\end{align}}
By substituting this expression in \eqref{E7}, we obtain
\begin{eqnarray}
&&\partial_{tt}\nu+4\left[\Omega^2+\frac{1}{2}
\left(\frac{\Delta g}{8\pi}\right)^2\nu^2\right]\nu=0, \label{duffinga}
\end{eqnarray}
which is recognized as the {\it Duffing equation} 
without the dumped and driven terms, having the 
{\it Jacobi elliptic functions} as exact 
solutions~\cite{2014Salas,2016Belendez}.
However, even before considering the explicit solution $\nu(t)$,
the exact period of oscillations can be obtained for \eqref{duffing}.
Eq.\eqref{duffing} also generalizes previous results given in Ref.~\cite{kenkre1987}, 
by including inter-species contributions in the interactions.
\vspace{-0.5cm}
\subsubsection{Period and amplitude of oscillations}
Multiplying \eqref{duffing} by $2\partial_t\nu$, we can obtain a 
time-invariant associated energy $E$, as
{\small\begin{eqnarray}
&&{\partial_t}\left\{
(\partial_t\nu)^2+4\Omega^2\nu^2+\left(\frac{\Delta g}{8\pi}\right)^2\nu^4\right\}=
2\frac{\partial E}{\partial t}=0, \nonumber\\
&&E=
\frac{1}{2}(\partial_t\nu)^2+2\Omega^2\nu^2+\frac{1}{2}\left(\frac{\Delta g}
{8\pi}\right)^2\nu^4.
\label{inv-E}
\end{eqnarray}
}As the energy is a constant, it can be obtained at the turning point (when
$\partial_t\nu=0$ and $\nu={\cal A}$), such that $E=2\Omega^2 {\cal A}^2+
\frac{1}{2}\left(\frac{\Delta g}{8\pi}\right)^2 {\cal A}^4$.
By integrating it within a time interval $\Delta T$ for which the 
atom-number difference  goes from $-{\cal A}$ to $+{\cal A}$, 
{\small\begin{eqnarray}
\int_0^{\Delta T} dt&=&\Delta T=\int_{-{\cal A}}^{{\cal A}}
\frac{d\nu}{\sqrt{
{{4\Omega^2}(A^2-\nu^2)-\left(\frac{\Delta g}{8\pi}\right)^2(A^4-\nu^4)}}}.
\nonumber\label{time-inv2}
\end{eqnarray}
}This interval is one-half of the density period, $\Delta T=T/2$.
With a variable transformation $\nu={\cal A} y$, 
we obtain 
the required relation between period $T$ and amplitude ${\cal A}$ through the known {\it 
Jacobi complete elliptic function of the first kind} $K(k)$ \cite{2021Whittaker}.
With $\alpha\equiv {\left(\frac{{\cal A}\,\Delta g}{8\pi}\right)}$, we have
{\small\begin{eqnarray}
T_K(\Omega,\alpha)&=&\int_{-1}^{1}\hspace{-0.1cm}
\frac{dy}{\sqrt{
{(1-y^2)\left[{\Omega^2}+\alpha^2(1+y^2)\right]}}}\nonumber\\&=&
\frac{2}{\sqrt{{\Omega^2}+\alpha^2}}K\left(\frac{-\alpha^2}{\Omega^2+\alpha^2}\right).
\label{period-ell2}
\end{eqnarray}
}Therefore, once given the parameters (in our case, the nonlinear interactions, Rabi constant and 
amplitude ${\cal A}$), we obtain the period. In the limit $\Delta g=0$, $T={\pi}/{\Omega}$, with
the other limit being for $\Omega\to 0$ (or $\alpha\gg \Omega$), where $K(-1)=1.311$.
However, one should notice that the period depends on the product
$(\Delta g {\cal A})^2$, instead of only on ${\cal A}^2$. 
With the conditions at $t=0$, where $\nu=0$ and $\partial_t \nu= 2\Omega$, 
and at the turning point, where $\nu={\cal A}$ and $\partial_t \nu=0$, we can obtain the
amplitude from the energy conservation:
{\small\begin{eqnarray}
\frac{(2\Omega)^2}{2}=2\Omega^2 {\cal A}^2+\frac{1}{2}\left(\frac{\Delta g}{8\pi}\right)^2 {\cal A}^4.   
\end{eqnarray}
}With $\alpha$ defined for \eqref{period-ell2}, and with $\alpha_0\equiv \frac{\Delta g}{8\pi}$, 
{\small\begin{eqnarray}
&&\alpha^4+(2\Omega)^2(\alpha^2-\alpha_0^2)=0,\;\;
\frac{\alpha^2}{2\Omega^2}=\sqrt{1+\left(\frac{\alpha_0}{\Omega}\right)^2}-1,\nonumber\\
&&|{\cal A}|={\sqrt{2}\frac{\Omega}{\alpha_0}}
\left[\sqrt{1+\left(\frac{\alpha_0}{\Omega}\right)^2}-1\right]^{1/2},\nonumber\\
&&|{\cal A}|_{\Omega\ll\alpha_0}\to \sqrt{\frac{2\Omega}{\alpha_0}},\;\; |{\cal A}|_{\Omega\gg\alpha_0}\to 1.
\end{eqnarray}
}By giving these conditions, the exact solution for $\nu(t)$ is also reachable, given by the Jacobi elliptic function, 
as shown in \cite{2014Salas}.
With the already given expressions for the period and amplitude, the exact solution for \eqref{duffing} can also be expressed in 
the sinusoidal form:
{\small\begin{eqnarray}
\nu(t)&=&{\cal A}\left[{1}-2\cos^2\left(t
\sqrt{\Omega^2+\frac{1}{2}\left(\frac{\Delta g}{8\pi}\right)^2{\nu_A^2}}
+\frac{\pi}{4}\right)\right]\nonumber\\
&=&{\cal A}\sin\left(2t
\sqrt{\Omega^2+\frac{1}{2}\left(\frac{\Delta g}{8\pi}\right)^2\nu_A^2}\right),
\label{dens-sol}
\end{eqnarray} 
}in which we assume as parameters $\nu_A$ and the amplitude 
${\cal A}$ of the density-difference oscillations, that are
closely related due to the periodic conditions.
As noticed from \eqref{dens-sol}, $\nu(t)$ also satisfies the harmonic oscillator
equation with time-dependent frequency, as the corresponding wave functions, 
but with one-half of the oscillating period. 
In our specific case, by solving it we obtain a relation between the amplitude ${\cal A}$ 
and the period $T$ for the oscillations of the atom-number ratio difference $\nu(t)$.
The agreement of \eqref{period-ell2} with the period $T$ obtained from \eqref{dens-sol} 
is shown in the lower panel of Fig.~\ref{fig01}.
\vspace{-0.5cm}
\subsubsection{Oscillating time-dependent phases}
The phase evolutions can be obtained by starting from  \eqref{nu1}  and \eqref{E10c}, 
considering $\psi_j={|\psi_j|}e^{-{\mathrm i}[\gamma_j t+\beta_j(t)]}$, as
{\small\begin{equation}
  \frac{1}{8\pi\Omega}\left(\frac{(\Delta g)\nu^2}{8\pi {\rm i}}+\partial_t\nu\right)=\psi_1^*\psi_2 .\label{E23} 
\end{equation}
}and  using the GP equation, to obtain 
{\small\begin{eqnarray}
\partial_t{\beta}_1&=&
\frac{\Delta g}{16\pi} {\nu} \frac{(2-{\nu})}{1-{\nu}},\;\;
\partial_t{\beta}_2=-
\frac{\Delta g}{16\pi} {\nu} \frac{(2+{\nu})}{1+{\nu}},\;\;
\label{E25}
\end{eqnarray}
}such that, with 
the explicit density-dependent part of the phases written as
{\small $\partial_t\beta_j=
\frac{\Delta g}{(8\pi)^2}\frac{1}{2 |\psi_j|^2}
\left[1-(8\pi)^2 |\psi_j|^4\right],$}
$\bar\gamma_j(t)=\gamma_j+\partial_t\beta_j$ can be written as
{\small\begin{eqnarray}
\bar\gamma_j(t)
&=&\gamma_0+
\left(\frac{\Delta g}{16\pi}\right)
\left[\frac{1}{8\pi |\psi_j|^2}-(8\pi |\psi_j|^2)\right].
\label{gamma2}
\end{eqnarray}
}\vspace{-0.5cm}

\section{Stability Methods Comparison}\label{app_comparison}

In section \ref{sec2}, it was already anticipated how the true
solutions can differ from stationary ones when the Rabi coupling $\Omega$ 
is leading the dynamics when the condensate wave functions are periodic
functions in time. In this section, we compare the BdG and Floquet 
spectra, which are actually suitable for stationary and periodic
time-dependent functions, respectively. As $\Omega$ increases, with  
the reduction of the population dynamics oscillating period, the
discrepancy between the approaches is more likely to increase. 

\begin{figure}[b]
\centering
    \includegraphics[scale=0.65]{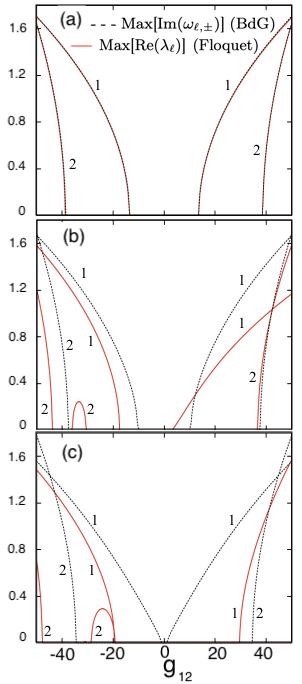}
    \vspace{-0.2cm}
\caption{(Color online) Maximum unstable spectra for  angular 
modes $\ell$ [$(=1,2)$, as indicated] are shown as functions 
of $g_{12}$, with fixed $g=1$ and frequencies $\Omega=$0.01, 0.50, and
0.99 [respectively, in panels (a), (b), and (c)]. 
With Max[Im($\omega_{\ell,\pm}$)] obtained from \eqref{bdg_spc}, the BdG
unstable spectra are shown with black-dashed lines. Floquet unstable spectra,
from \eqref{floquet_eq}, provided by Max[Re($\lambda_{\ell}$)], are 
with red-solid lines. Within defined units, all quantities are presented as dimensionless.}
\label{fig13}
\end{figure}
In Fig.~\ref{fig13}, we set together all stability approximations previously
discussed in order to observe more closely how the approaches are related 
to each other. 
In panels \ref{fig13}{ (a)}, \ref{fig13}{ (b)} and 
\ref{fig13}{ (c)} we display three different regimes of Rabi
coupling. In the weak regime $\Omega=0.01$, we see both, BdG and Floquet,
approaches return about the same spectrum as already expected.
Conversely, when the Rabi coupling constant is increased, the spectrum 
of the analytical approaches becomes quite different from each other.

\begin{figure}[t]
    \centering  
\includegraphics[scale=0.4]{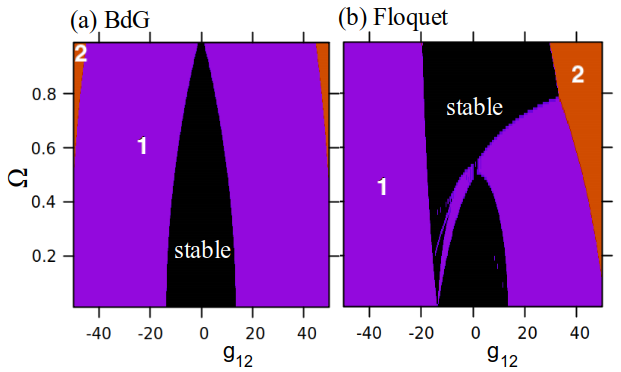} 
\vspace{-0.2cm}
\caption{(Color online) 
Stability diagrams of Rabi parameter $\Omega$ versus $g_{12}$ for fixed $g=1$, 
given by  Max[Im($\omega_{\ell,\pm}$)], representing stable regions ($\le 0$, Black regions), and 
unstable ones [Violet, for $\ell=1$ and orange, for $\ell=2$].
BdG approach is in (a) [see~\eqref{bdg_spc}], with  Floquet approach in (b) [see~\eqref{floquet_exp}].
Within defined units, all quantities are presented as dimensionless.}
\label{fig14}
\end{figure}

\begin{figure}[h]
    \centering
    \includegraphics[scale=0.4]{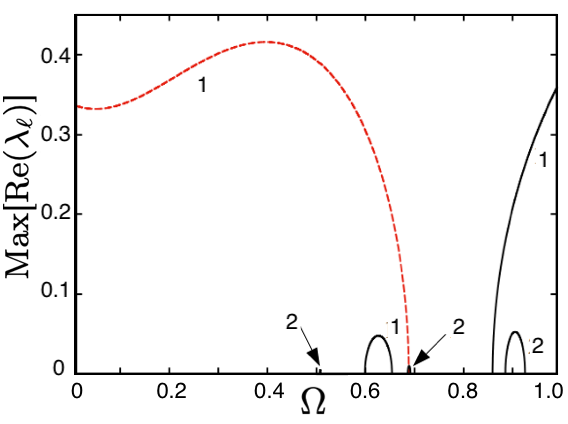}\vspace{-0.2cm}
\caption{(Color online) Floquet maximum spectra (Max[Re($\lambda_{\ell}$)]) of  
unstable angular modes $\ell=1,2$ (indicated by the numbers)
as functions of the 
coupling $\Omega$
[see~\eqref{floquet_eq}]. Cases ($g$, $g_{12}$) $=$ (1, 15) and (40, -10) are displayed with red-dashed  
and black-solid lines, respectively. 
Within defined units, all quantities are presented as dimensionless.}\label{fig15}
\end{figure}

A deeper analysis is provided in  Fig.~\ref{fig14}, 
which displays simultaneously the different roles of the Rabi coupling
depending on the inter-species interaction. It is very clear that the 
coupling is able to open a large region of stability, which makes the 
unstable behavior to be postponed. But in some situations, it can make 
the system unstable, even when the BdG spectrum provides no trace of
instability.
This phenomenon is also depicted for fixed parameters in Fig.~\ref{fig15}. 
We display the two different sets of interaction parameters ($g$, $g_{12}$)
$=$ (1, 15) and (40, -10), the first one is driven from an unstable to a
stable solution by the increasing of the Rabi coupling, and the second one 
is led from a stable regime to an unstable one when the coupling becomes
higher.
{\small\begin{table}[h]
\caption{Dynamics stability status of the BEC mixture by the three methods, 
given the parameter of intra- and inter-species interaction, and Rabi coupling constant 
$\Omega$, $g$, and $g_{12}$, respectively. Unstable cases are displayed with the dominant unstable mode $\ell$. The reference figures where the results can be checked are set in parentheses.}
\label{tab1}
\centering
    \begin{tabular}{c c c  c c c}
    \hline\hline    
$\Omega$ & $g$ & $g_{12}$ &  BdG & Floquet & Dynamics \\ \hline \hline
 0.50 & 1 & 8 & stable  & $\ell=1$  & $\ell=1$   \\ 
& & & (\ref{fig13}{b}) & (\ref{fig13}{b}) & (\ref{fig07}, \ref{fig08}{a}, \ref{fig10})  \\ \hline
 0.94 & 40 & -10 & stable  & $\ell=1$ & $\ell=1$    \\
 & & &  (\ref{fig03}{b} ) &  (\ref{fig15}) & (\ref{fig07}, \ref{fig08}{b}, \ref{fig09}{a}, \ref{fig11}{a-c})
\\ \hline
 0.10 & 1 & 10 & stable  & stable  & stable   \\
 & & & (\ref{fig03}{a}) & (\ref{fig04}{a}) & (\ref{fig07})
\\ \hline
 0.99 & 1 & 25 & $\ell=1$  & stable  & stable   \\
 & & & (\ref{fig13}{c}) & (\ref{fig13}{c}) & (\ref{fig07})
\\ \hline
 0.90 & -10 & 20 & $\ell=1$  & $\ell=2$  & $\ell=2$   \\
 & & & (\ref{fig03}{b}) & (\ref{fig04}{b}) & (\ref{fig09}{b}, \ref{fig11}{e-f})
\\
\hline\hline
\end{tabular}
\vspace{-0.5cm}
\end{table}}
In table \ref{tab1}, we compare the predictions of BdG and Floquet methods with the full 
dynamics calculations for the five sets of parameters ($\Omega$, $g$, $g_{12}$) mostly discussed 
in the text. We can check that the Floquet 
spectrum agrees with the dynamics simulations for all cases, then it is more suitable 
for our system than the BdG method.
\vspace{-.5cm}
\section{Resonance conditions} 
\label{app_resonance_v2}

In the regime $g_{12} \ll g$ (or, equivalently, $\Delta g \ll g$ ), 
let us consider $\psi_{j}(\theta,\phi,t)=\phi_j(t){\mathrm e}^{-{\mathrm i}(\mu+\delta\mu)t},$ 
where $\mu=(2g+\Delta g)/(8\pi)$ is the chemical potential, with a $\Delta g$- first- order phase correction $\delta\mu=-\Delta g/(16\pi)$. The functions $\phi_j(t)$ given by
$\phi_{1}=({1}/{\sqrt{4\pi}}\cos\left(\Omega t+\frac{\pi}{4}\right)\left[1+{\rm i}\Delta(t)\right]$, 
$\phi_{2}=({1}/{\sqrt{4\pi}}\sin\left(\Omega t+\frac{\pi}{4}\right)\left[1-{\rm i}\Delta(t)\right]$
where we are taking into account a time-dependent $\Delta g$- 
first-order correction function of interaction difference $\Delta g $, given by 
$\Delta(t)=\frac{\Delta g}{32\pi\Omega}\cos(2\Omega t)$. 
Small amplitude fluctuations around $\psi_j(\theta,\phi,t)$ are assumed, as
\begin{eqnarray}
    \psi_j(\theta,\phi,t)&=&[\phi_j(t)+\delta\phi_j(\theta,\phi,t)].
    \label{eq_app_pert_func}
\end{eqnarray}
Now, we apply the following useful transformation from $(\delta\phi_1,\delta\phi_2)$ 
to $(\delta\phi_d,\delta\phi_s)$~\cite{2022Zhang}, in which terms higher than first order
$\Delta(t)$ are neglected:
\begin{subequations}
\begin{eqnarray}
    \begin{pmatrix}
     \delta\phi_d \\
     \delta\phi_s
    \end{pmatrix}
    &=&
    \begin{pmatrix}
        {\phi}_{1}^{*} && {\phi}_{2}^{*} \\
        -{\phi}_{2} && {\phi}_{1}
    \end{pmatrix}
    \begin{pmatrix}
     \delta\phi_1 \\
     \delta\phi_2
    \end{pmatrix}, \\
    \begin{pmatrix}
     \delta\phi_1 \\
     \delta\phi_2
    \end{pmatrix}
    &=&4\pi
    \begin{pmatrix}
        \phi_1 && -{\phi}_{2}^{*} \\
        {\phi}_{2} && {\phi}_{1}^{*} 
    \end{pmatrix}
    \begin{pmatrix}
     \delta\phi_d \\
     \delta\phi_s
    \end{pmatrix}.\end{eqnarray}
\end{subequations}
With \eqref{eq_app_pert_func} and the above inserted in \eqref{eq01}
(neglecting 2nd and higher-order terms in $\Delta g$, $\delta\phi_d$, and $\delta\phi_s$), 
the following coupled equation for the excitations is obtained:
{\small
\begin{eqnarray}
{\mathrm i}\frac{\partial\delta\phi_d}{\partial t}&=&\Big\{-\frac{1}{2}\nabla^2 +  
\frac{g}{4\pi}+\frac{\Delta g}{16\pi}\left[1+2\cos(4\Omega t)\right] \Big\}\delta\phi_{d}\nonumber \\
&+&\Big\{\frac{g}{4\pi}+\frac{\Delta g}{16\pi}\left[1+\cos(4\Omega t) \right] \Big\}\delta{\phi}_{d}^{*}
\nonumber \\
&+&\Big\{\frac{\Delta g-2g}{8\pi}\delta{\phi}_{s} -\frac{\Delta g+3g}{16\pi}
    \delta{\phi}_{s}^{*}\Big\} \sin(4\Omega t),\\
    {\mathrm i}\frac{\partial\delta\phi_s}{\partial t}&=&
    \Big\{-\frac{1}{2}\nabla^2 + \frac{\Delta g}{16\pi}\left[1-2\cos(4\Omega t)\right] \Big\}\delta\phi_{s}
    \nonumber \\
    &-&\Big\{\frac{\Delta g}{16\pi}\left[1+\cos(4\Omega t)\right] \Big\}\delta{\phi}_{s}^{*}\nonumber\\
        &-&\Big\{\frac{\Delta g}{8\pi}\delta{\phi}_{d}
    +\frac{\Delta g}{16\pi}\delta{\phi}_{d}^{*}\Big\}\sin(4\Omega t),
\end{eqnarray}
}leading to a BdG matrix determinant for the spectrum, $|{\bf Z}-\omega_\ell|=0$,
where, in 1st-order of $\Delta g$ and dropping the sinusoidal terms~\cite{2019Chen} 
(nullified in the respective periods)
{\small
\begin{eqnarray} {\bf Z}=
\begin{pmatrix}
\epsilon_\ell+\frac{4g+\Delta g}{16\pi} & \frac{4g+\Delta g}{16\pi} & 0 & 0 \\
-\frac{4g+\Delta g}{16\pi} & -\epsilon_\ell -\frac{4g+\Delta g}{16\pi}& 0 & 0 \\
0 & 0 & \epsilon_\ell+\frac{\Delta g}{16\pi}& -\frac{\Delta g}{16\pi} \\
0 & 0 & \frac{\Delta g}{16\pi} & -\epsilon_\ell-\frac{\Delta g}{16\pi}
\end{pmatrix}.
\end{eqnarray}
}The solution leads to the natural elementary excitations:
{\small
    \begin{eqnarray}
        \omega_{d}^2={\epsilon_\ell\left(\epsilon_\ell+\frac{g}{2\pi}+\frac{\Delta g}{8\pi}\right)}, 
        \;{\rm and}\;
        \omega_{s}^2 ={\epsilon_\ell\left(\epsilon_\ell+\frac{\Delta g}{8\pi}\right)}.
\label{C6}    \end{eqnarray} 
}
  
\end{document}